\documentclass[preprint]{elsarticle}
\usepackage{epstopdf}
\usepackage{caption}
\usepackage{tikz}
\usepackage{overpic}
\usepackage{import}
\usepackage{here} 
\usepackage{pgf}
\usepackage[utf8]{inputenc}
\usepackage[switch,modulo]{lineno}

\usepackage{booktabs}
\usepackage{enumitem}
\usepackage{geometry}

\usepackage[version=3]{mhchem}
\usepackage{mathrsfs}

\usepackage{multicol}
\usepackage{nomencl}
\usepackage{etoolbox}

\usepackage{hyperref}

\usepackage{graphicx}
\usepackage{booktabs}
\usepackage{multirow}
\usepackage{microtype}
\usepackage{longtable}
\usepackage{tabularx}
\usepackage{doi}
\usepackage{dblfloatfix}
\usepackage{soul}
\usepackage{siunitx}
\usepackage{framed}
\usepackage{nomencl}
\usepackage{derivative}
\usepackage[export]{adjustbox}
\usepackage{float}
\usepackage[T1]{fontenc}
\usepackage{xcolor}
\usepackage{tcolorbox}\tcbuselibrary{breakable}
\usepackage{amsmath}
\usepackage{amssymb}
\usepackage{soul,xcolor}
\usepackage{subfigure}
\journal{International Journal of Hydrogen Energy}

\newcommand{\ep}[1]{{\color{black}{#1}}}
\newcommand{\dmm}[1]{{\color{black}{#1}}}
\newcommand{\tg}[1]{{\color{black}{#1}}}
\newcommand{\emf}[1]{{\color{black}{#1}}}
\newcommand{\dmmrev}[1]{{\color{black}{#1}}}

\begin{document}
\setstcolor{red}

\begin{frontmatter}

\title{\LARGE Analysis of thermodiffusive instabilities in hydrogen premixed flames using a tabulated flamelet model}

\cortext[cor1]{Corresponding author: D. Mira, daniel.mira@bsc.es}
\author{
    { 
    Emiliano M. Fortes$^{a}$, Eduardo J. Pérez-Sánchez$^{a}$, Ambrus Both$^{a}$,\\
    Temistocle Grenga$^{b}$, Daniel Mira$^{a,*}$}\\[10pt]
    {\footnotesize \em $^a$Barcelona Supercomputing Center (BSC), Plaça Eusebi Güell, 1-3 08034, Barcelona, Spain}\\[-5pt]
    {\footnotesize \em $^b$Faculty of Engineering and Physical Sciences, University of Southampton, SO17 1BJ, Southampton, United Kingdom}\\[-5pt]
}

\date{}

\begin{abstract}

Preferential diffusion effects play a paramount role in the evolution of lean premixed hydrogen flames since they directly impact flame surface corrugation, which, in turn, affects the flame behaviour from a macroscopic point of view. Simulating such flames with tabulated chemistry (TC) methods remains challenging due to difficulties in capturing the full complexity of the flame dynamics. A comprehensive characterization of the capabilities and limitations of flamelet-based manifolds to reproduce these dynamics is still needed.
In this work, a comprehensive formulation including preferential diffusion effects through mixture-averaged molecular diffusion in the context of tabulated chemistry is applied to the study of the propagation and structure of freely propagating hydrogen flames where intrinsic instabilities play an important role.
The performance of the tabulated approach is evaluated by comparing its predictions with detailed chemistry (DC) calculations. The analysis focuses on two key aspects: the model behaviour in linear and non-linear 
regimes and the sensitivity of the response of the model to pressure and temperature variations.
Additionally, the impact of the mesh resolution on the flame response is examined in order to determine the capabilities of the proposed method 
in the absence of subgrid models.
The analysis begins by examining the linear regime through the dispersion relation.
The results show that the thermodynamic conditions may significantly impact the range of wave numbers susceptible of being well-predicted by the tabulated model: the increase of either temperature or pressure, corresponding to more realistic engine operating conditions, can noticeably extend such range. However, some divergences of the dispersion relation in the linear regime, especially for the stable range, are found and show certain tendency of the tabulated model to slight overpredict the flame wrinkling. 
Subsequently, the non-linear regime is analysed by computing global flame parameters and comparing the flame structure with the reference solutions.
The results show that the model can capture global flame descriptors accurately for the three conditions investigated with relative errors of less than 20$\%$. Considering the complexity of the physical and chemical phenomena involved, it can be concluded that the model successfully reproduces the most relevant effects governing flames exhibiting thermodiffusive instabilities and offers a reliable alternative to DC with notably lower computational cost.

\end{abstract}


\end{frontmatter}

\section{Introduction}

Lean premixed hydrogen flames have gained increased attention and have become a topic of significant importance because of their potential role in sustainable energy conversion systems. However, these flames exhibit unconventional characteristics when compared to traditional hydrocarbon fuels, necessitating a deeper understanding and the development of specific closure models. Of particular interest, the flame front in these flames features intrinsic instabilities, such as Darrieus-Landau~\cite{matalon2007intrinsic, Matalon2009} and thermodiffusive instabilities. The latter, caused by significant differences in molecular and thermal diffusivities also referred as Lewis number variations, can play a fundamental role in practical applications due to enhancements in flame speed \cite{Berger2022IntrinsicInstabilitiesPremixed_I}. 
In particular, in hydrogen flames, thermodiffusive instabilities primarily arise from the low Lewis number of the fuel. This disparity leads to a self-excited wrinkling of the flame front, characterised by the increase of flame surface and the formation and destruction of cellular-like structures for lean equivalence ratios below a critical value.
The increase in flame surface area is associated with fluctuations in the reaction rate due to variations in the local equivalence ratio, which in turn enhance flame reactivity and increase the global burning velocity~\cite{Berger2022IntrinsicInstabilitiesPremixed_II}.
Furthermore, the strength of these instabilities is dependent on the equivalence ratio, temperature, and pressure of the unburnt mixture~\cite{Berger2022IntrinsicInstabilitiesPremixed_I}, highlighting the need for models that can reproduce these effects across a wide range of conditions~\cite{LAPENNA2024113126}.
Therefore, \ep{it is required to extend well-established combustion models to capture such effects.}  \ep{A wide perspective of the particular characteristics of hydrogen combustion and the different approaches devised in the literature to incorporate the thermodiffusive instabilites in the modelling is given in \cite{Pitsch2024}.} 
In particular, developing reduced-order models or manifold representations that can include both the chemical effect and thermodiffusive behaviour of these flames is of paramount importance for the development of next-generation hydrogen-based combustion technologies. 

Over the years, the community has focused on developing several formulations for reduced-order models based on the flamelet hypothesis. In this context, the Flamelet-Generated Manifold method (FGM)~\cite{VANOIJEN201630}, the Flame Prolongation of Intrinsic low-dimensional manifolds (FPI)~\cite{Gicquel2000} and the Flamelet Progress Variable (FPV)~\cite{Knudsen2009} have been successfully applied in various configurations. 
These models assume that the thermochemical states of the flame essentially remain on a manifold dependent on a given set of variables known as controlling variables. Usually, the manifold is constructed from one-dimensional flames~\cite{VANOIJEN201630} due to its simplicity and reduced computational cost. Incorporation of thermodiffusive effects in the manifolds is not straightforward, and the extension of the models to include such effects is an active area of research. Unlike conditions with unity Lewis number, the diffusion coefficient of each species shows a different ratio with respect to the thermal diffusion.
In turn, under certain conditions, this thermodiffusive process leads to more wrinkled fronts that can strongly alter the development of the flame.
Specifically, this phenomenon enhances flame corrugation when it has a destabilising effect, significantly affecting the flame speed~\cite{Berger2022IntrinsicInstabilitiesPremixed_I}.
Capturing these effects through tabulated methods is essential to recover the flame behaviour.

In the FGM context, a first approach devised in~\cite{Vreman2009} considers effective Lewis numbers for each control variable of the manifold. This approach was later refined by introducing new terms in the control variable transport equations, which comprised the crossed effect between control variables arising from the preferential diffusion of some species~\cite{deSwart2010, Donini2015}. Such terms were defined based on the gradients in phase space leading to coefficients tabulated in the manifold. This approach has been applied to the simulation of premixed flames for hydrogen/methane mixtures and, more recently, to Large Eddy Simulations (LES)~\cite{Almutairi2023}.
Based on physical arguments, all these works argued that the thermodiffusive process was locally a function of the progress variable only and, therefore, only some of the cross terms were accounted for.
Conversely,~\cite{Abtahizadeh2015, Nicolai2022} presented a complete formulation considering the cross terms under the constant non-unity Lewis numbers hypothesis. A different strategy for detailed transport with constant Lewis numbers was recently proposed in the context of FGM~\cite{Mukundakumar2021} by grouping the contribution of the non-unity Lewis numbers into new coefficients.
However, despite the model's improved predictions from previous models~\cite{Donini2015}, its extension to mixture-averaged transport is not straightforward.

Regarding the FPV model, pioneer work to include preferential diffusion effects was introduced in~\cite{Regele2013}, where only the mixture fraction transport equation was modified by including a source term depending on the progress variable. While in this original work only the Lewis number for the fuel was considered different to unity, the model was later extended in~\cite{Schlup2019} to incorporate a fully mixture-averaged approximation and thermal diffusion effects. A new variant to tackle preferential diffusion in the frame of FPV through the mixture-averaged model was presented in~\cite{Bottler2022}, where species mass fractions were transported using a mixture-averaged diffusion model and then used to reconstruct the control variables in order to access the manifold. This approach was applied to the simulation of several configurations for hydrogen-air mixtures. Also, an FPV formulation that included curvature and strain was tested in~\cite{Wen2022_part_I} to investigate ultra-lean premixed hydrogen flames with significant thermodiffusive instabilities at atmospheric and high pressure. Both models were compared in~\cite{Bottler2023} to simulate expanding spherical flames. This study found that despite
the model can have some impact on the prediction of the wrinklimg associated with small wavelengths, the overall prediction of the flame behaviour was satisfactorily captured. However, gaps still exist in understanding the predictive capabilities of tabulated chemistry models across a wide range of operating conditions~\cite{Berger2022DevelopmentOfLargeEddySimulations}.

In general, it can be said that models reproduce the global behaviour of the flame, although predicting the effects in the range of the small wavelengths is challenging \cite{Bottler2023}.
As demonstrated in~\cite{matalon2007intrinsic}, this range is influenced by non-unity Lewis number effects and capturing its behaviour through tabulated methods is particularly complex~\cite{Bottler2023}.
Understanding the impact of such small wavelengths on the flame through global quantities (flame speed, flame surface area) remains open. Therefore, it is interesting to determine how relevant the deviations introduced by the model in such a range of wavelengths affect its capabilities to predict the global flame characteristics.

To further examine the ability of tabulated methods to capture intrinsic instabilities in lean hydrogen premixed flames, a formulation incorporating mixture-averaged diffusion transport is employed~\cite{FGM-Mix}. This formulation accounts for the contributions of all cross terms, as detailed in Sec.~\ref{subsec:methodology}.
For consistency, contributions related to molecular weight and velocity correction are also incorporated into the present formulation. This tabulated flamelet model offers a robust description of preferential diffusion at a low computational cost. In a previous study by the authors~\cite{FGM-Mix}, the formulation details were thoroughly explained, along with the key differences from the previously discussed methods. Furthermore, the model was applied for a systematic evaluation of a set of canonical configurations for stratified flames. The previous study concluded that the model accurately captures the flame structure and propagation speed in lean hydrogen flames subjected to spatial variations in mixture fraction.

The aim of this paper is, therefore, to critically evaluate the results and limitations of the proposed combustion model in order to ensure a comprehensive understanding of its applicability to lean hydrogen flames where strong intrinsic instabilities appear.
Both the linear and non-linear regimes are analysed to provide an integral description
of the model response and comparisons with detailed chemistry solutions are used to delimit the capabilities of the tabulated approach. The method is applied to various conditions featuring different strengths of preferential diffusion effects in order to evaluate the generality and applicability of the proposed tabulated chemistry method for general applications in premixed combustion.
Additionally, the study aims to determine the spatial resolution requirements to obtain the global descriptors of the flame during the non-linear regime when no sub-grid modelling is included, through the examination of meshes with varying resolutions.

The paper is organised as follows. Section~\ref{sec:methodology} details the model and presents the computational cases. Section~\ref{sec:results} provides a detailed analysis of the results, examining both the linear and non-linear regimes. Finally, Section~\ref{sec: conclusions} presents the conclusions drawn from the study and suggests directions for future work.

\section{Methodology\label{sec:methodology}}

In this work, a tabulated chemistry (TC) method based on a database of laminar premixed flamelets is extended to incorporate preferential diffusion effects through the mixture-averaged diffusion model~\cite{FGM-Mix}. The solutions are compared with those obtained using detailed chemistry (DC) to assess the predictive capabilities of the proposed approach in freely propagating lean hydrogen flames exhibiting strong thermodiffusive effects.
Details of the transport equations, tabulation strategy and numerical solver are given below.

\subsection{Theoretical description\label{subsec:methodology}} \addvspace{10pt}

Detailed simulations including preferential diffusion effects require an accurate description of the diffusive transport, mainly determined by the local variations in Lewis number. This work considers mixture-averaged transport approximation using a velocity correction to ensure mass conservation without Soret and Duffour effects. While these effects are relevant for the accurate description of the fundamental properties of the flame, these terms are neglected in this study to focus exclusively on the ability of the proposed tabulated flamelet model to recover the diffusive fluxes with preferential diffusion.
The mixture-averaged approximation, also referred as Hirschfelder and Curtis approximation~\cite{hirschfelder1964molecular}, provides the best first-order approximation to the solution of the exact diffusive velocities, and gives the best trade-off between computational cost and accuracy to describe preferential diffusion effects. Numerical simulations with DC and TC using the same numerical methods and fluid solver are used in order to ensure a fair comparison between the two approaches. 

For both approaches, the low Mach number approximation of the Navier-Stokes equations~\cite{Both2020LowdissipationFiniteElement} is solved, which leads to the following continuity and momentum equations:
\begin{equation}\label{eq:continuity}
    \frac{\partial \rho}{\partial t} + \mathbf{\nabla} \cdot \left(\rho \mathbf{u} \right) = 0,
\end{equation}
\begin{equation}\label{eq:navier-stokes}
    \frac{\partial \left(\rho \mathbf{u}\right)}{\partial t} + \mathbf{\nabla} \cdot \left(\rho \mathbf{u}  \otimes \mathbf{u}\right) = - \mathbf{\nabla}p + \mathbf{\nabla} \cdot \mathbf{\tau}.
\end{equation}

Here, $\rho$ denotes the mixture density, $\mathbf{u}$ the velocity vector, $p$ the pressure, and $\mathbf{\tau}$ the shear stress tensor. The combustion process is described by the governing equations for species mass fractions $Y_k$ (with $k$ going from $1$ to the number of species $N_s$), which are given by:

\begin{equation}\label{eq:mass-fractions}
    \frac{\partial \left(\rho Y_{k} \right)}{\partial t} + \mathbf{\nabla} \cdot \left(\rho \mathbf{u} Y_{k}\right) + \mathbf{\nabla}\cdot \mathbf{j}_{k}  = \dot{\omega}_{k},
\end{equation}

\noindent where $\mathbf{j}_{k}$ represents the diffusive flux of the $k$-th species and $\dot{\omega}_{k}$ its chemical source term. 
Using a mixture-averaged transport model, the diffusive flux $\mathbf{j}_{k}$ of species k is given by:

\begin{equation}\label{eq:species-diffusive-flux}
    \mathbf{j}_{k} = \rho \mathbf{V}_{k} Y_{k} = -\rho D_{k}\frac{W_{k}}{W} \mathbf{\nabla} X_{k} + \rho \mathbf{V}^{c} Y_{k}.
\end{equation}

Equation~(\ref{eq:species-diffusive-flux}) contains
the diffusive velocity $\mathbf{V}_{k}$, the mixture-averaged diffusion coefficient $D_{k}$, the molecular weight of the $k$-th species $W_{k}$, the molecular weight of the mixture $W$ and the molar fraction $X_k$.

A correction velocity $\mathbf{V}^{c}$ is introduced in the mixture-averaged model to ensure mass conservation, \ep{that is,} $\sum_{k=1}^{N_{s}} Y_k \mathbf{V}_{k} = 0$, yielding $\mathbf{V}^{c} = \sum_{j=1}^{N_{s}} D_{j}\frac{W}{W_{j}} \mathbf{\nabla} X_{j} $. The diffusion coefficient of the $k$-th species is determined by the binary diffusion coefficients $\mathcal{D}_{jk}$ through the following equation~\cite{hirschfelder1964molecular, Chapman1999MathematicalTheoryOfNonUniformGases}:
\begin{equation}
    D_{k} = \frac{1-Y_k}{\sum_{\substack{j=1 \\ j \neq k}}^{N_s} X_j / \mathcal{D}_{j k}}.
\end{equation}

Finally, the enthalpy equation can be expressed with the same notation as for the mass species mass fractions and leads to:

\begin{equation}\label{eq:enthalpy}
    \frac{\partial \left(\rho h \right)}{\partial t} + \mathbf{\nabla} \cdot \left(\rho \mathbf{u} h\right) + \mathbf{\nabla}\cdot \mathbf{j}_{h}  = 0,
\end{equation}
where the diffusive flux of enthalpy $\mathbf{j}_{h}$ is given by:

\begin{equation}\label{eq:enthalpy-diffusive-flux}
    \mathbf{j}_{h}  = - \lambda \mathbf{\nabla} T + \sum_{k=1}^{N_{s}} \rho \mathbf{V}_{k} Y_{k} h_{k},
\end{equation}
and $\lambda$ is the thermal conductivity, $T$ is the temperature and $h_k$ is the enthalpy for the $k$-th species.

The proposed tabulated chemistry model~\cite{FGM-Mix}, referred here as TC, is based on pre-computed, one-dimensional adiabatic laminar premixed flames that cover a representative range of mixture fractions within the flammability limits. Due to the compositional variations caused by the preferential diffusion of certain species, at least two coordinates are necessary to parametrise the thermochemical states~\cite{Berger2022IntrinsicInstabilitiesPremixed_II}. In this study, the mixture fraction, $Z$, and a chemical progress variable, $Y_c$, are utilised. The resolution of thermochemical quantities in DC versus TC manifolds and laminar configurations has been extensively characterised \cite{Mukundakumar2021,FGM-Mix}.

The chemical evolution from unburnt to burnt conditions is described by a progress variable $Y_{c}$, defined as a linear combination of species mass fractions $Y_{c} = \sum_{k=1}^{N_{s}} \alpha_{k} Y_{k}$, where $\alpha_k$ are constants chosen to define a monotonic evolution of $Y_{c}$ with respect to the flame spatial coordinate, allowing for a reparametrization of the thermochemical states as a function of $Y_c$.
The transport equation for $Y_{c}$ is obtained from adding the species mass fractions weighed by coefficientes $\alpha_k$: 
\begin{equation}\label{eq:reactive-progress-variable}    
\frac{\partial \left(\rho Y_{c} \right)}{\partial t} + \mathbf{\nabla} \cdot \left(\rho \mathbf{u} Y_{c}\right) + \mathbf{\nabla}\cdot \mathbf{j}_{Y_c}  = \dot{\omega}_{Y_c},
\end{equation}
\noindent where both the diffusive and source terms are defined as linear combinations of individual quantities, $\mathbf{j}_{Y_c} = \sum_{k=1}^{N_{s}} \alpha_{k} \mathbf{j}_{k}$ and $\dot{\omega}_{Y_{c}} = \sum_{k=1}^{N_{s}} \alpha_{k} \dot{\omega}_{k}$.

The fuel/air mixture is represented by Bilger's mixture fraction $Z$.
Given an element $p$, the elemental mass fraction of the element $Z_{p}$ can be defined as:
\begin{equation}\label{eq:elemental-mass-fraction}
Z_{p} = W_{p} \sum_{k=1}^{N_s} a_{kp} \frac{Y_k}{W_p},
\end{equation}
where $a_{kp}$ denotes the number of atoms of element $p$ in the composition of the $k$-th species.
From this quantity, Bilger's mixture fraction is defined as a linear combination of elemental mass fractions of the involved elements. Considering hydrogen as fuel (non-carbon fuel), the mixture fraction $Z$ can be obtained as: 

\begin{equation}\label{eq:mixture-fraction-bilger}
Z=\frac{\frac{Z_{H}-Z_{H,2}}{2W_{H}}-\frac{Z_{O}-Z_{O,2}}{W_{O}}}{\frac{Z_{H,1}-Z_{H,2}}{2W_{H}}-\frac{Z_{O,1}-Z_{O,2}}{W_{O}}}=K_Z\left(\frac{Z_{H}-Z_{H,2}}{2W_{H}}-\frac{Z_{O}-Z_{O,2}}{W_{O}}\right),
\end{equation}
\noindent where the inverse of the denominator of the first equality is denoted for simplicity as $K_{Z}$ and the number subscripts referring to $1$ and $2$ denote the fuel (H$_2$) and oxidiser (air) streams, respectively. Finally, by linearly combining the equations from \eqref{eq:mass-fractions} and \eqref{eq:elemental-mass-fraction},
the transport equation for Bilger's mixture fraction is obtained:
\begin{equation}\label{eq:bilger-mixture-fraction}
\frac{\partial (\rho Z)}{\partial t}+ \nabla \cdot (\rho \boldsymbol{u} Z) + \nabla \cdot(\mathbf{j}_{Z})=0,
\end{equation}
\noindent where the diffusive flux for mixture fraction $\mathbf{j}_{Z}$ is the linear combination of the atomic diffusive fluxes $\mathbf{j}_{Z_{p}}$, given by:
\begin{equation}\label{eq:dif_flux_Z}
\mathbf{j}_{Z}=K_Z \left(\frac{\mathbf{j}_{z_{H}}}{2 W_H}- \frac{\mathbf{j}_{z_{O}}}{W_{O_2}}\right).
\end{equation}
In turn, the atomic diffusive fluxes $\mathbf{j}_{Z_{p}}$ are also defined by linear combinations of the species mass fraction diffusive fluxes \eqref{eq:species-diffusive-flux}, so a compact form for the element diffusive fluxes can be obtained:
\begin{equation}\label{eq:dif_flux_Zp}
\mathbf{j}_{z_{\mathbf{p}}}=W_p \sum_{k=1}^{N_s} \frac{a_{k p}}{W_k} \mathbf{j}_k.
\end{equation}

The dependence generated in the manifold ($\psi=\psi(Z,Y_c)$ for any thermochemical quantity $\psi$) allows to express the diffusive fluxes for species $\mathbf{j}_k$, where the gradients of the species appear, as function of the gradients of $Z$ and $Y_c$ after applying the chain rule:

\begin{equation}\label{eq:chain-rule}
    \mathbf{j}_{k} = - \rho D_{k}\frac{W}{W_k} \left(\frac{\partial X_k}{\partial Y_c} \mathbf{\nabla} Y_{c} + \frac{\partial X_k}{\partial Z} \mathbf{\nabla} Z\right) + \sum_{j=1}^{N_{s}} \rho D_{j}\frac{W}{W_j} Y_{k} \left(\frac{\partial X_j}{\partial Y_c} \mathbf{\nabla} Y_{c} + \frac{\partial X_j}{\partial Z} \mathbf{\nabla} Z\right).
\end{equation}

Then, the combination of the species diffusive fluxes results in the diffusive fluxes for the progress variable and mixture fraction, according to $\mathbf{j}_{Y_c} = \sum_{k=1}^{N_{s}} \alpha_{k} \mathbf{j}_{k}$ and equations \eqref{eq:dif_flux_Z} and \eqref{eq:dif_flux_Zp}, respectively. This allows the fluxes to be rewritten as the sum of some coefficients times the gradient of the controlling variables. Such coefficients, denoted here as $\{\Gamma_{Y_{c}, Y_{c}}, \Gamma_{Y_{c}, Z}, \Gamma_{Z, Y_{c}}, \Gamma_{Z, Z}\}$, depend on the pair $(Z,Y_c)$ and can be precomputed and stored in the flame database directly from the set of laminar flames.
Details about their calculation are given in \cite{FGM-Mix}.

Replacing Eq.~\eqref{eq:chain-rule} in the transport equations for the progress variable (Eq.~\eqref{eq:reactive-progress-variable}) and Bilger's mixture fraction (Eq.~\eqref{eq:bilger-mixture-fraction}), and grouping all the terms 
for $\nabla Y_c$ and $\nabla Z$, yields the final transport equations used in the proposed TC method:
\begin{equation}\label{eq:fgm_Yc}
\rho \frac{\partial Y_c}{\partial t} + \rho \mathbf{u} \cdot \nabla Y_c = \nabla \cdot (\rho \Gamma_{Y_c,Y_c} \, \nabla Y_c + \rho \Gamma_{Y_c,Z} \, \nabla Z ) + \rho \dot{\omega}_{Y_c},
\end{equation}
\begin{equation}\label{eq:fgm_Z}
\rho \frac{\partial Z}{\partial t} + \rho \mathbf{u} \cdot \nabla Z = \nabla \cdot (\rho \Gamma_{Z,Y_c} \, \nabla Y_c  + \rho \Gamma_{Z,Z} \, \nabla Z ),
\end{equation}

The effects of preferential diffusion are {encapsulated} in the coefficients $\{\Gamma_{Y_{c}, Y_{c}}, \Gamma_{Y_{c}, Z}, \Gamma_{Z, Y_{c}}, \Gamma_{Z, Z}\}$.
Note that these equations are exact, and the only assumption made is the flamelet concept. It is worth mentioning that the same procedure used to define the coefficients $\{\Gamma_{Y_{c}, Y_{c}}, \Gamma_{Y_{c}, Z}, \Gamma_{Z, Y_{c}}, \Gamma_{Z, Z}\}$ can be extended to any other variable, including enthalpy. The derivation of analogue $\Gamma$ coefficients for the enthalpy equation is omitted here because it is not required for the adiabatic simulations of this paper. Additional details of its derivation can be found in the original paper by the authors~\cite{FGM-Mix}.

\subsection{Numerical framework\label{subsec:methodology1}} \addvspace{10pt}

The numerical simulations using DC and TC were conducted with the parallel multiphysics code Alya developed at the Barcelona Supercomputing Center~\cite{Vazquez2016AlyaMultiphysicsEngineering}. Alya employs a second-order spatial scheme using linear finite elements. For the current simulations, a low-Mach number approximationfor the Navier-Stokes equations using a low-dissipation fractional step method is used \cite{Both2020LowdissipationFiniteElement}. An explicit third-order Runge-Kutta scheme is applied for temporal integration of momentum and scalars. The DC model has been validated in previous works~\cite{MIRA20235091,KALBHOR2023112868,RAMIREZMIRANDA2023105723,SURAPANENI2023112715}, while the TC model described in~\cite{FGM-Mix} has been integrated in the tabulated chemistry framework from the Alya code~\cite{Mira10.1115/GT2018-76229, Mira2020NumericalCharacterization, BENAJES2022111730}.
The chemistry of this work uses a detailed reaction mechanism for $\mathrm{H_2}$/air combustion from Burke et al.~\cite{Burke2012ComprehensiveKineticModel} containing nine species with 19 reactions. As previously mentioned, the mixing dimension is characterised by $Z$ \eqref{eq:bilger-mixture-fraction} while the progress variable \eqref{eq:reactive-progress-variable} is defined as water vapour mass fraction, $Y_c=Y_{\mathrm{H_2O}}$, because of its monotonic increasing behaviour across the flame front in the one-dimensional flames. This choice has previously shown excellent results for lean hydrogen mixtures~ \cite{Berger2022IntrinsicInstabilitiesPremixed_II} and is retained here. Data retrieval from the manifold in the TC model is performed using a second-order interpolation algorithm, while the tables are generated in a representative range of mixture fractions which span from $Z_{min}=0$ to $Z_{max}=0.054$ using the interpolation proposed in our previous work~\cite{FGM-Mix}. The manifold space is built with a normalised 
progress variable defined as $c=(Y_c-Y_{c,u}(Z))/(Y_{c,b}(Z)-Y_{c,u}(Z)$, where subscripts $u$ and $b$ refer to the unburnt and burnt gases, respectively. For the sake of clarity, the mixture fraction dependence of $Y_{c,u}$ and $Y_{c,b}$ has been made explicit in this expression. The calculation of the one-dimensional flames used to construct the combustion databases have been done using the solver Cantera~\cite{cantera}.

It is worth mentioning that the proposed tabulated chemistry model leverages substantial reductions in computational cost. For the simulations in this paper, an average speedup factor of 10x is obtained with TC with respect to DC using Alya in the MareNostrum IV supercomputer.

%
%
\subsection{Simulation setup\label{subsec:methodology2}} \addvspace{10pt}

\begin{figure}
\centering
\includegraphics[]{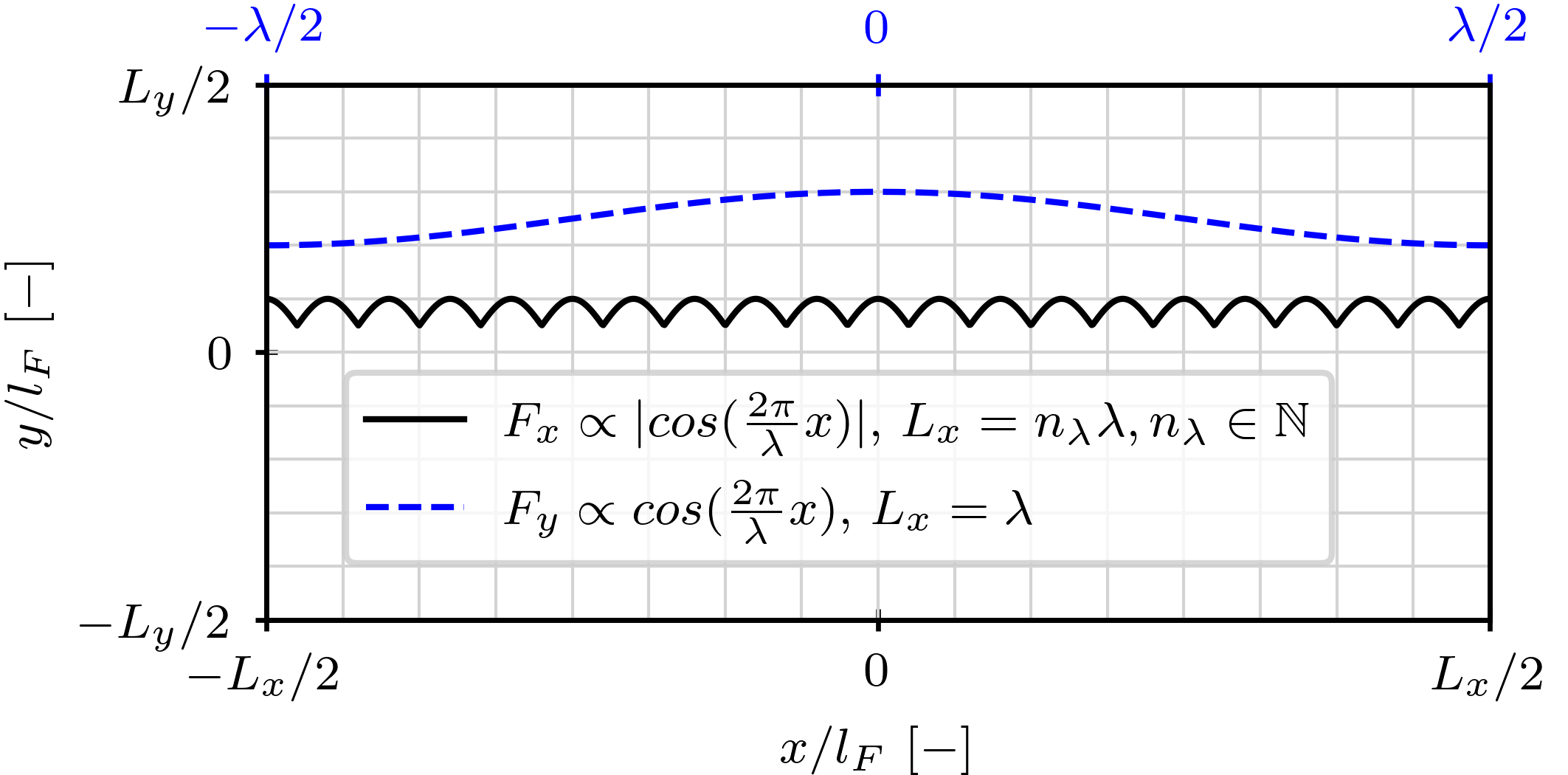}
\caption{\footnotesize Schematic of the initial conditions used for the premixed flames.
The solid black line represents the initial condition for the non-linear regime analysis, while the blue dashed line corresponds to the one of the linear regime.
}\vspace{-2.5mm}
\label{fig: schematic}
\end{figure}

The physical problem corresponds to the study of freely propagating two-dimensional premixed flames at different operating conditions. The problem is solved in a rectangular domain of dimensions $L_{x}$ and $L_{y}$ using a uniform Cartesian mesh with elements of regular size.
A premixed $H_2$-air mixture is injected from the bottom of the domain ($y=-L_y/2$) at an equivalence ratio $\phi=0.5$ with a constant velocity $u_{in}$. Periodic boundary conditions are applied along the left- and right-hand sides, while a zero Neumann condition is imposed at the outlet, located at $y=L_{y}/2$ (see Fig. \ref{fig: schematic}).

To ensure a proper resolution of the flame front, the mesh spacing is chosen based on the one-dimensional unstretched adiabatic flame thickness for the inlet equivalence ratio, defined as 

\begin{equation}\label{eq:thermal_flame_thickness}
    l_F=\frac{T_b-T_u}{\mathrm{max}(|\nabla T|)},
\end{equation}

\noindent where $T_u$ and $T_b$ are the temperatures for the unburnt and burnt gases, respectively. 
Then, the mesh spacing, denoted by $\Delta$, is computed as a fraction of the flame thickness $l_F$, that is, $\Delta = l^{in}_F/n$, being $n$ an integer. 
Additionally from $l^{in}_F$, a characteristic time scale $\tau=l^{in}_F/s^{in}_L$ is defined, being $s^{in}_L$ the flame speed for such unstretched flame.

For each condition two resolutions are considered, a fine mesh with $n=10$ and a coarse one with $n=5$. The results obtained with the fine mesh are considered fully converged and serve as a reference solution. In contrast, the second mesh is included to show the capabilities of the TC model for coarse resolutions (the DC with $n=5$ is only included for completeness), as it usually occurs in situations where the resolution is low e.g. LES. Comparisons between the cases with $n=10$ with $n=5$ are included 
to quantify the influence of mesh resolution on the predictive capabilities of the model.

Finally, the initial condition is imposed using the solution of the one-dimensional unstretched flame at the equivalence ratio $\phi=0.5$ along the $y$ direction. To promote the generation of instabilities the flame front is perturbed by a periodic function $F(x)$. The choice of such function depends on the type of analysis (linear or non-linear) and is detailed in the following subsections. 
The configuration of the problem is depicted in Fig.~\ref{fig: schematic}.

The cases under investigation include three operating conditions of practical interest, going from atmospheric to elevated pressure and temperature.
These thermochemical states are based on previous studies~\cite{Berger2022IntrinsicInstabilitiesPremixed_I, Berger2022IntrinsicInstabilitiesPremixed_II} and summarized in Table~\ref{tab:main_cases}. 
Case I is the baseline case and corresponds to a mixture with low unburnt temperature $T_u$ and atmospheric pressure, conditions introducing strong effects of thermodiffusive instabilities. In Case II, a preheating of the mixture is considered, attenuating the hydrodynamic and thermodiffusive effects (decrease of thermal expansion ratio and Zeldovich number). 
Finally, Case III features a pressure increase of factor 5x, further enhancing thermodiffusive effects (increase of the Zeldovich number) and burning velocities. The different cases are solved for both DC and TC models using the two mesh resolutions given above ($n=5, 10$).

\begin{table}
\centering
\begin{center}
\begin{tabular}{|c|c|c|c|c|c|c|c|c|}
\hline
Case  & $p$ [atm] & $T_u$ [K] & $u_{in}/s_{L}$ & $l_{F}$ $[\mu m]$ & $s_{L}$ $[m/s]$ & $\tau$ $[m/
s]$ & $L_{x} / l_{F}$ & $L_{y} / l_{F}$ \\ \hline
I & 1 & 298 & 1.4 & 423 & 0.49 & 0.85 & 200 & 200 \\ 
II & 1 & 700 & 1 & 507 & 5.30 & 0.095 & 100 & 150 \\ 
III & 5 & 298 & 1.8 & 118 & 0.22 & 0.53 & 100 & 150 \\ 
\hline
\end{tabular}
\end{center}
\caption{Inlet conditions (columns 2 to 4), flame properties of the corresponding unstretched one-dimensional flames (columns 5 to 7), and domain dimensions for the non-linear regime simulations (columns 8 and 9).}
\vspace{-2mm}
\label{tab:main_cases}
\end{table}

%
%
\subsubsection{Linear regime setup}\label{subsubsec: methodology-linear}

Perturbing a planar flame with a weak harmonic signal triggers a \dmmrev{flame response with an}
amplification or attenuation of the initial perturbation amplitude \ep{depending on the wavelength}. 
The response is characterized by an exponential growth or decay in the perturbation amplitude when sufficiently weak harmonic disturbances are applied.
Large wavelengths (small wave numbers) often lead to destabilization due to the dominance of hydrodynamic instabilities (Darrieus-Landau instability), whereas small wavelengths (high wave numbers) tend to stabilize the flame due to the energy and mass flux balances when curvature and strain rate effects are non-negligible.

To analyze the initial growth or decay of small perturbations, the flame front is perturbed according to the function: 
\begin{equation}\label{eq:linear-regime-perturbation}
F(x) = A_{0} \, \mathrm{cos}\left(kx\right),
\end{equation}
where $A_{0}$ is the initial amplitude of the perturbation and $k$ is the wave number, which determines the wavelength of the perturbation. 
The perturbation wavelengths $\lambda = \frac{2 \pi}{k}$ are defined as multiples of $l_F$, while the domain length in the $x$ direction is chosen to be a multiple of or equal to the perturbation wavelength $\lambda$. 
This ensures that a full period of the perturbation fits within the physical domain and can be seen in blue color in Fig.~\ref{fig: schematic}. For all simulations, the height of the domain is set to a sufficiently large value of $L_{y}=30l_{F}$ to prevent any influence from the outlet boundary. The initial flame is positioned at the mid-height of the domain, with an initial amplitude for the perturbation of $A_{0}=0.04l_F$. Moreover, the inlet velocity is set as the unstretched laminar flame speed obtained using Cantera \cite{cantera}. Columns 5 to 7 from table~\ref{tab:main_cases} provide additional details of the one-dimensional flame properties for the three inlet conditions.

The evolution of the perturbation on the flame front is tracked following the amplitude along the temperature isoline $\mathcal{C}_{T_{val}}$ for a fixed temperature value $T_{val}$. Initially, the amplitude and shape of the perturbation follow the selected function \eqref{eq:linear-regime-perturbation}. 
The amplitude of the flame front may have an exponential growth (positive or negative)
($A(t)\propto e^{\omega t}$) up to a critical time $t_{crit}$.
For $t<t_{crit}$, period of time for which the flame evolution can be considered in the linear regime, the perturbed flame front is spatially well-described by a sinusoidal function. In this range, an essentially constant amplitude can be found according to

\begin{equation}\label{eq:growth_rate}
\omega = \frac{d \mathrm{ln} \left( A(t) \right)}{dt}.
\end{equation}

\noindent By computing the response $\omega$ at different wavelengths, the dispersion relation $\lambda$ is calculated.

On the contrary, beyond this critical time, which delimits the linear regime, the flame front shape deviates significantly from a sinusoidal function and the flame front profile becomes more complex showing contributions from wavelengths different to the one of excitement (non-linear regime).

\subsubsection{Non-linear regime setup}\label{subsubsec: methodology-non-linear}

For the non-linear regime analysis, in order to rapidly promote perturbations in a rich variety of wavelengths (to reduce the computational cost), a perturbation function that introduces a wide set of wave numbers is used: 
\begin{equation}\label{eq:non-linear-perturbation}
    F(x) = A_0 \left| \mathrm{cos}\left( kx \right) \right|.
\end{equation}

\noindent The promotion of instabilities is achieved by using an absolute value function. The Fourier series of this function shows null value for the sinus coefficients and only non-null for the cosine coefficients related to $\lambda/(2m)$ with $m$ being a natural number and $\lambda$ the fundamental wavelength. In particular, such coefficients are in the form $(-1)^{m+1}$ $4/((4m^2-1) \pi))$. If carefully choosing the fundamental wavelength based on the dispersion relations, as there exists a wide set of wavelengths contained in this function, there will be content in the unstable range that will trigger the instability in a short amount of time. All the simulations have been performed with a fundamental wavelength of $\lambda = 4 l_F$. 
To allow an analysis of the flame with no influence of the boundary conditions, the dimension of the domain $L_{x}$ is chosen large enough to accommodate several periods of the perturbation function, and $L_{y}$ is chosen to ensure that the flame has sufficient time to develop instabilities inside the domain.
Columns 7 to 9 from Table ~\ref{tab:main_cases} show the conditions and domain dimensions of the three cases considered in this study.

%
%
\section{Results and discussion}{\label{sec:results}}

Numerical simulations using the tabulated chemistry TC model are compared with the detailed chemistry DC solutions for the three operating conditions given in
Table ~\ref{tab:main_cases}.
The first subsection ~\ref{subsec: results} focuses mainly on the assessment of the TC approach 
\dmmrev{in the non-linear regime}
to predict the flame structure and shape, including the thermodiffusive fluxes. The second subsection~\ref{subsec:results2} addresses \dmmrev{the linear regime
through the study of the dispersion relations for the proposed flamelet model,} while the third subsection~\ref{subsec:results3} is focused on \dmmrev{the prediction of macroscopic flame parameters in the non-linear regime.}


\subsection{Flame structure and thermodiffusive flux analysis}{\label{subsec: results}}

\begin{figure*}
\centering
\vspace{-0.4 in}
\includegraphics[width=0.96\linewidth]{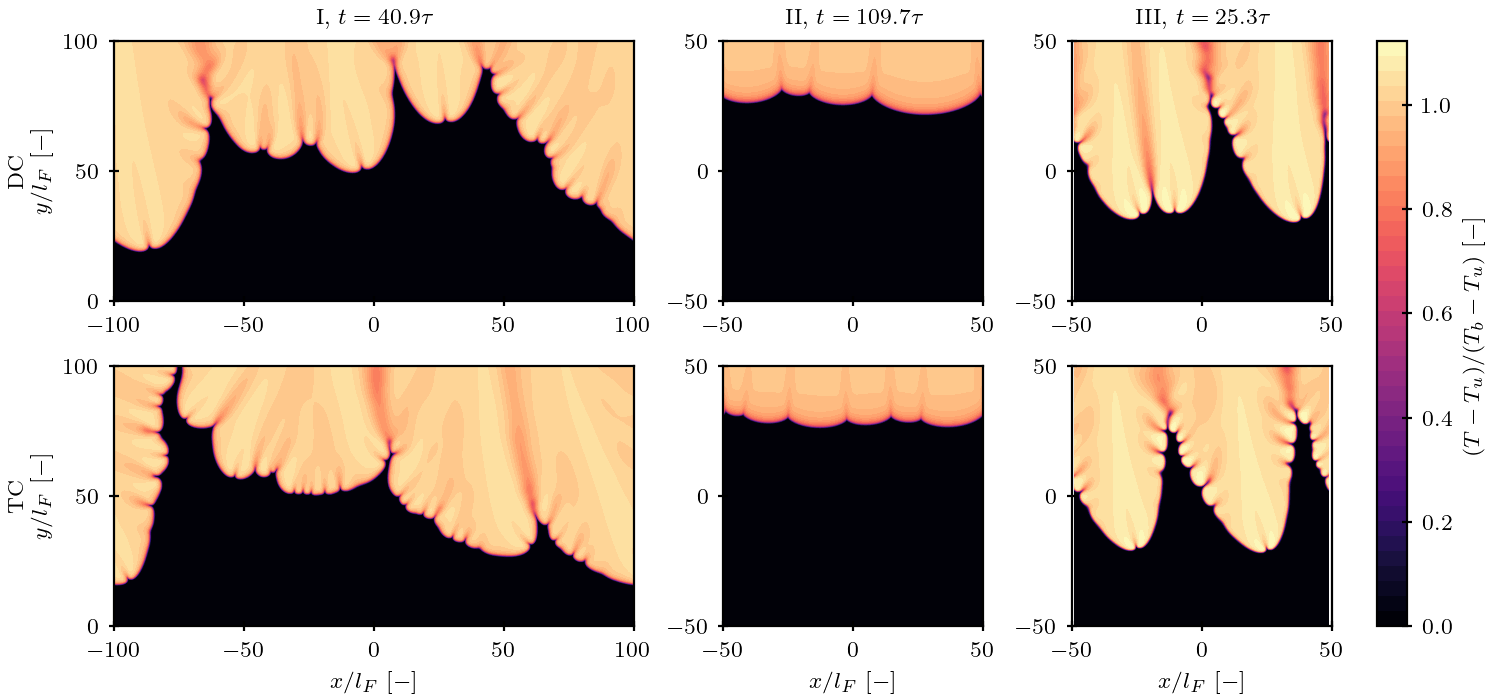}
\vspace{10 pt}
\caption{\footnotesize Contour plots of normalized temperature with respect to the adiabatic temperature of the unstretched one-dimensional flame for the three conditions of Table~\ref{tab:main_cases} for each model with $n=10$: DC (first row) and TC (second row). The adiabatic temperatures are $T_{b}=1638.2$K (Case I), $T_{b}=1973.3$K (Case II) and $T_{b}=1639.8$K (Case III). The time step is indicated as a multiple of the flame characteristic time $\tau$. Plots show only a subregion of the domain for better visualization. }\vspace{-2.5mm}
\label{fig: contours}
\end{figure*}

A direct comparison of the fully developed flame topology (non-linear regime) obtained with TC and DC is shown in Fig.~\ref{fig: contours} for the three cases using the fine mesh ($n=10$).
The flames feature distinctive small-scale cellular structures and wrinkling patterns when going to high temperature or pressure, as already evidenced in the literature ~\cite{Berger2022IntrinsicInstabilitiesPremixed_II}. The proposed TC approach captures these characteristic flame structures qualitatively, replicating their number, size, and curvature with remarkable accuracy. In particular, the TC method predicts the formation of large-scale cusps for the high-temperature (Case II) while reproducing the small-scale cellular structures and the characteristic sharp indentations pointing towards the burnt gases of the thermodiffusively unstable conditions (Cases I and III)~\cite{Matalon2018,BERGER20231525}. Also, the long trails formed downstream of the flame with super- and sub-adiabatic temperatures at the convex and concave regions \ep{(relative to the fresh gases)} of the flame front, respectively, are well-predicted by the TC model. However, the model predicts a more wrinkled flame front compared to the DC due to the presence of smaller wavelengths that distort the flame sheet. \dmmrev{Further discussion about this point is given in the next subsection.}


\begin{figure*}
\centering
\includegraphics[width=0.99\linewidth]{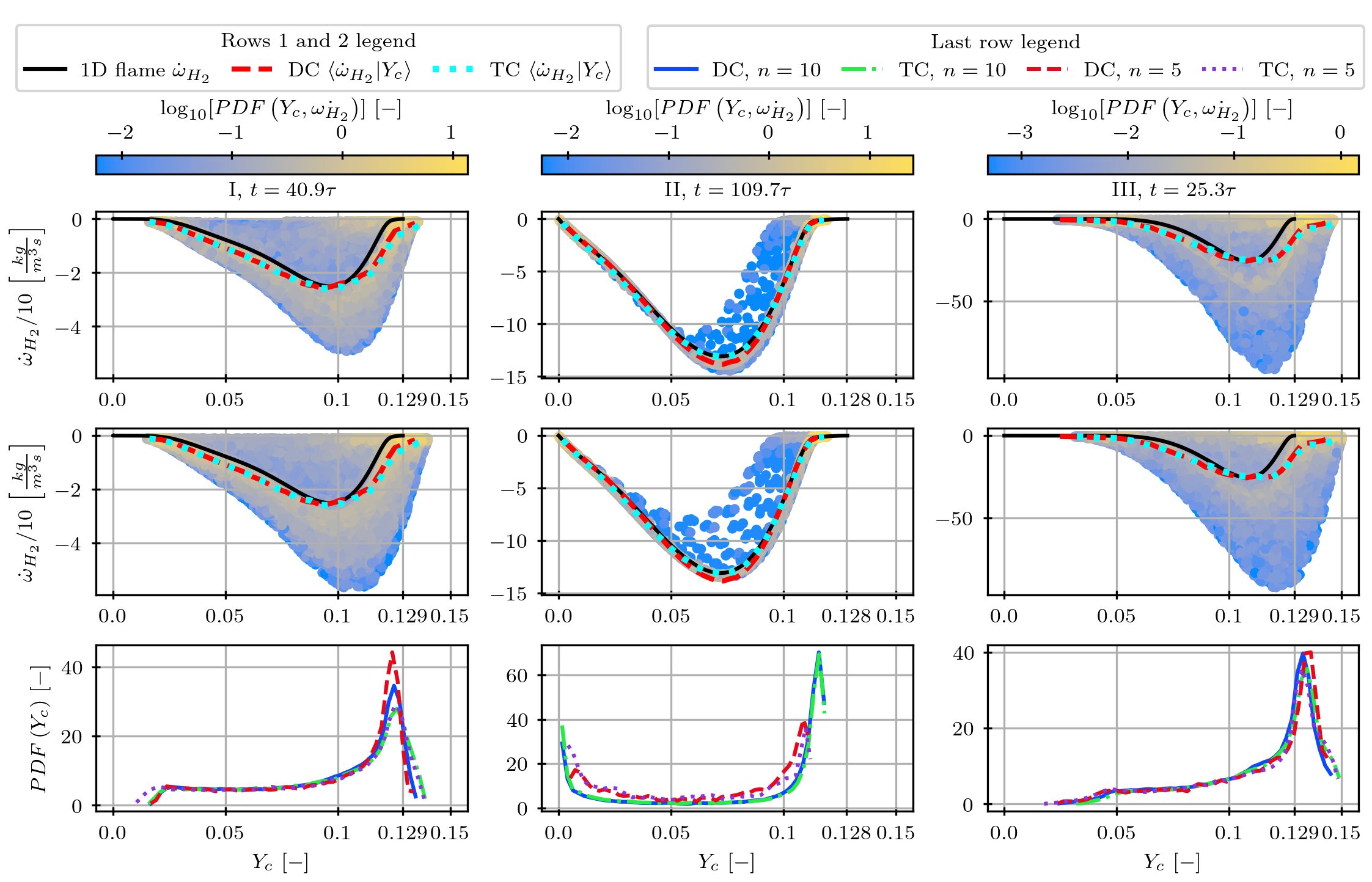}
\caption{\footnotesize \textbf{First and second rows}: each panel presents scatter plots of $\dot{\omega}_{\mathrm{H}_2}$ vs $Y_{c}$ coloured by the $\mathrm{log}_{10}$ values of the jPDFs $PDF \left( Y_{c}, \dot{\omega}_{H_{2}}\right)$ for flames I (left column), II (middle column) and III (right column) for the models DC (top row) and TC (middle row) with mesh resolution $n=10$ for the time instants shown in Fig.~\ref{fig: contours}. Also, the source term for the one-dimensional unstretched flame for the inlet mixture fraction (black solid lines) and $\langle \dot{\omega}_{\mathrm{H_2}} | Y_c \rangle$ for each model (red dashed lines for DC and cyan dotted lines for TC) with $n=10$ are included for reference. \textbf{Bottom row}: marginal PDFs for $Y_c$ for DC with $n=10$ (blue solid lines),  DC with $n=5$ (red dashed lines), TC with $n=10$ (green dash and dotted lines) and TC with $n=5$ (purple dotted lines). To only include representative states of the flame, only physical points where $\dot{\omega}_{H_2} < \epsilon < 0$ are included (with $\epsilon=-1 \mathrm{kg/(m^3 s)}$). Maximum values of $Y_{c}$ for the unstreteched one-dimensional flames are included in the axis to ease the visualization.}
\label{fig: histograms}
\end{figure*}

A quantitative analysis of the flame structure is shown in 
Fig.~\ref{fig: histograms} by the analysis of probability density functions. The top and middle rows present the joint probability density functions (jPDFs) of the net production rate 
of hydrogen $\dot{\omega}_{\mathrm{H_2}}$ and $Y_c$, denoted as $PDF(Y_c,\dot{\omega}_{\mathrm{H_2}})$, \ep{for the DC (top row) and TC (middle row) once the \dmmrev{perturbed flame front} is fully developed (non-linear regime). For each operating condition, \dmmrev{jPDFs 
for both models at the same time instant are presented.} Additionally,} the unstretched one-dimensional flame at the inlet equivalence ratio and the conditional average source term to the progress variable, $\langle  \dot{\omega}_{\mathrm{H_2}} | Y_c \rangle$, are included  \dmmrev{in Fig.~\ref{fig: histograms} as reference}. 
Note that the null values of $\dot{\omega}_{H2}$ have been excluded from the analysis to prevent skewing the PDFs towards the fully burnt and unburnt mixture states.

The scatter plots reveal a clear correlation between the source term distribution and the progress variable for all cases. Starting from zero source term \dmmrev{on the unburnt side}, the flame reaches a peak reactivity zone characterized by \dmmrev{a maximum \ep{in absolute value of } 
$\dot{\omega}_{H2}$ \ep{(minimum $\dot{\omega}_{H2}$ since it is negative)} that, finally, recovers the equilibrium state reaching the value of zero on the burnt side.}
The area covered by the scatter plots exhibits an almost identical match between $PDF(Y_c,\dot{\omega}_{\mathrm{H_2}})$ for the DC (top row) and TC (middle row) simulations for the three conditions tested, with only 
\ep{limited} differences in the peak values of $\dot{\omega}_{H2}$ for the reference Case I. 
\dmmrev{This highlights the capabilities of the TC model at representing the characteristics of flames in the presence of thermodiffusive instabilities.}

\ep{The representations for Cases I and III (first and third columns)} reveal a \emf{significant scatter in} the reaction rates \dmmrev{across the flame front}. These effects correspond to reacting fronts 
\emf{characterised} by high wrinkling and small-scale structures (Fig.~\ref{fig: contours}) \ep{where} local variations in mixture fraction \ep{are present} due to preferential diffusion \ep{effects}. Although scattering around the one-dimensional profile is present along the whole evolution in $Y_c$, it intensifies at intermediate stages of evolution and near the burnt side, where the flame composition departs more noticeably from the unstretched one-dimensional flame structure. 
\dmmrev{In contrast, Case II remains constrained to the mean value since the flame front is less curved and thus less prone to deviate from the unstretched one-dimensional flame structure due to the weaker effect of thermodiffusive instabilities, see Fig.~\ref{fig: contours}.}


\emf{In particular,} local enrichment of the mixture fraction (with respect to the inlet value) is found at the convex regions of the flame front \ep{(\emf{relative to} the fresh gases)}, while pockets with leaner mixture fractions are formed along concave zones. \ep{Cases I and III, for which the preferential diffusion effects 
\emf{lead to signficant deviations} from the one-dimensional flame structure,
show an enhanced source term (in absolute value) with high probability (points below the one-dimensional profile) due to the local enrichment in agreement with Berger et al.~\cite{Berger2022IntrinsicInstabilitiesPremixed_II}.}
On the other hand, a low reactivity branch is also detected around $\dot{\omega}_{\mathrm{H_2}} \approx 0$ \dmmrev{for leaner conditions, which leads to an overlap of the conditional average source term $\langle\dot{\omega}_{H_2} | Y_c\rangle$ with that of the unstretched one-dimensional flame. The results show the TC model is able to capture this behaviour.}


In contrast, the increase of inlet temperature (Case II) weakens the formation of thermodiffusive instabilities and a more tightly distributed jPDF is formed. As a result, the scattering around the mean value reduces and the flame structure of the unstretched one-dimensional flame is \ep{essentially} recovered.
\ep{As a consequence,} it can be inferred that at least two controlling variables are necessary to properly describe the internal structure of a flame with thermodiffusive instabilities \cite{Berger2022IntrinsicInstabilitiesPremixed_II}.
Additionally, an increase in the Zeldovich number, either by raising the pressure or lowering the temperature, leads to a shift in the jPDF distribution towards higher values of $c$, which is well captured by the flamelet model.

\ep{Finally,} the marginal PDFs of $Y_c$ for \ep{both} TC and DC with the two resolution levels \dmmrev{are shown in the bottom row of Fig.~\ref{fig: histograms}}. Such PDF consistently shows a peak around high $c$ values, close to equilibrium, reflecting the presence of the long trails that extend downstream the flame already observed in Fig.~\ref{fig: contours} for the temperature field.
Moreover, the profiles indicate a good correlation between the TC data \dmmrev{with} the DC solution at both refinement levels, ensuring a correct prediction of the flame structure for all three cases. \ep{A systematic analysis has been conducted for radical species, such as H, which plays a significant role in preferential diffusion, \emf{as well as} O and OH, for which similar representations are presented in \ref{appendix:radicals}. Excellent agreement is once again observed, demonstrating that the TC model provides reliable predictions of the flame structure.}


To further investigate the predictive capabilities of the TC method, the diffusive fluxes for H$_2$ are analysed. For this purpose, the balance of these fluxes in the normal and tangential directions is examined at an early stage ($t\approx1.4\tau$ for flames I and III and $t\approx5\tau$ for flame II) and during stable propagation (see Fig.~\ref{fig: contours} \dmmrev{for reference} times).
These balances are calculated using the normal unit vector $\mathbf{\hat{n}}$, defined as $\nabla c / |\nabla c|$, and the tangential vector, $\mathbf{\hat{t}}$, which is orthogonal to $\mathbf{\hat{n}}$ and defined by the components $t_x=-n_y$ and $t_y = n_x$ \ep{(90$^{\circ}$ counterclockwise rotation)}. Referring to the flux for hydrogen as $\mathbf{j}_{\mathrm{H}_2}$, its local projection into the normal and tangential directions to the flame front, $\mathbf{j}_{\mathrm{H}_2}= (\mathbf{j}_{\mathrm{H}_2} \cdot \mathbf{\hat{n}}) \mathbf{\hat{n}} + (\mathbf{j}_{\mathrm{H}_2} \cdot \mathbf{\hat{t}}) \mathbf{\hat{t}}$, facilitates the computation of the divergence of the normal and tangential fluxes as follows:

\begin{equation}\label{eq:div_normal}
    \nabla \cdot ((\mathbf{j}_{\mathrm{H}_2} \cdot \mathbf{\hat{n}}) \mathbf{\hat{n}}),
\end{equation}

\begin{equation}\label{eq:div_tan}
    \nabla \cdot ((\mathbf{j}_{\mathrm{H}_2} \cdot \mathbf{\hat{t}}) \mathbf{\hat{t}}).
\end{equation}

\begin{figure*}
\centering
\includegraphics[width=0.99\linewidth]{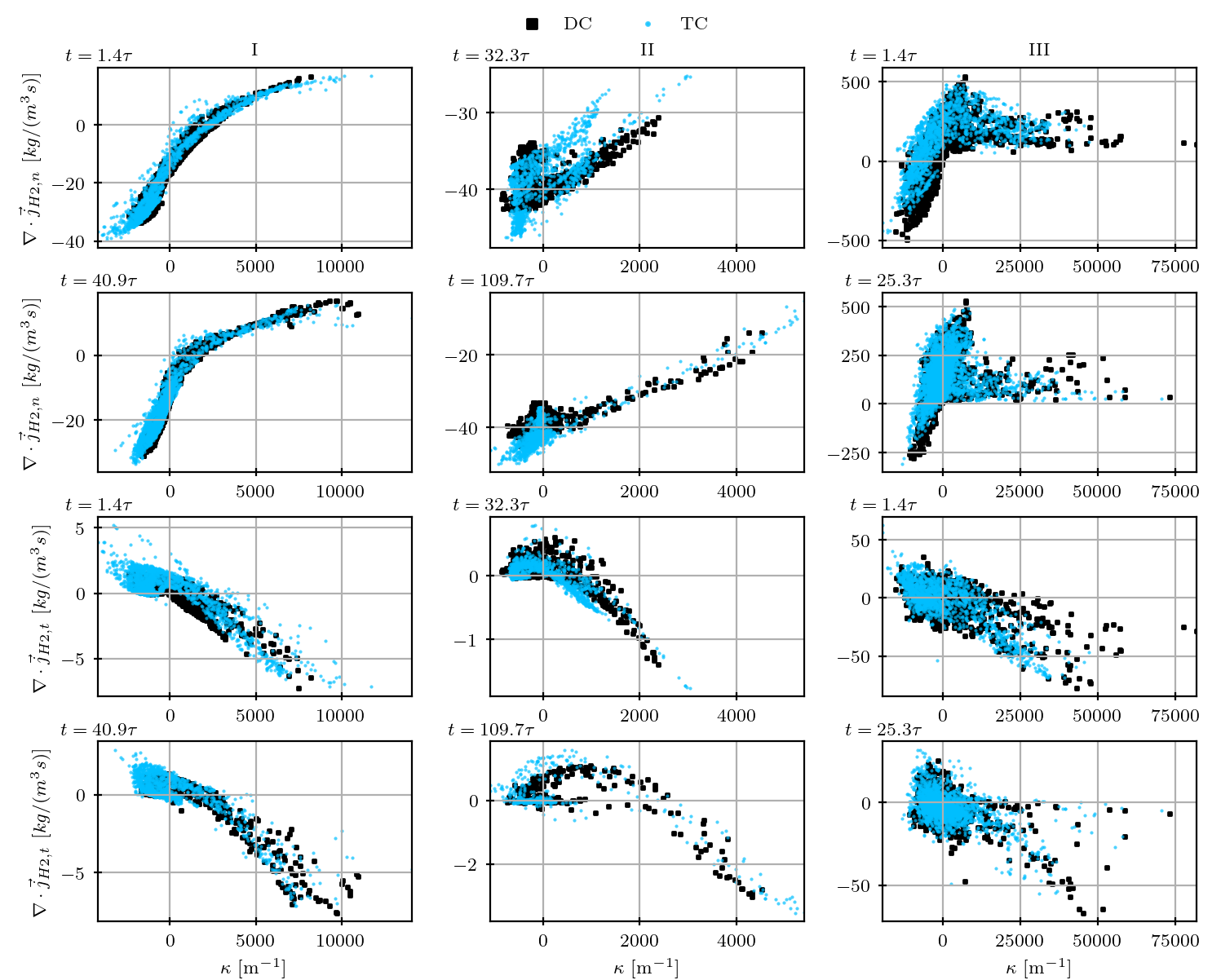}
\caption{\footnotesize \emf{Scatter plots showing the normal and tangential contributions of H$_2$ diffusive fluxes versus flame front curvature along the iso-line $c=$ 0.8, for the three cases described in Table \ref{tab:main_cases}. Each column represents a different flame simulation: Case I (left), Case II (center), and Case III (right). The first two rows depict the normal direction diffusion at an early time (first row) and a later time (second row). The third and fourth rows show the tangential direction diffusion for the same time points. In each panel, scatter plots for models with a resolution of $n=10$ are displayed, with the DC model represented by black squares and the TC model by blue circles.}
\label{fig: diffusive fluxes}}

\end{figure*}

The fluxes are evaluated as a function of curvature, $\kappa=\vec{\nabla}\cdot \hat{\mathbf{n}}$, along the \ep{$c=$ 0.8} 
isoline, as shown in Fig.~\ref{fig: diffusive fluxes}. \ep{For both components,} the highest absolute curvatures are found in the concave regions of the flame front \ep{(seen from the fresh gases)}, corresponding to intrusions amplified by preferential diffusion (Le < 1), where the maximum diffusive fluxes are also observed. In contrast, the convex regions, where the flame front is less curved, exhibit more moderate flux values, closer to those of an unstretched one-dimensional flame. Additionally, \dmmrev{the point density in the scatter plot is lower for the concave regions than for the convex ones, where the flux behaviour is more uniform. In fact, the normal fluxes are substantially larger than the tangential ones, even under conditions of curvature and stretch.
In the convex regions, the normal flux is about one order of magnitude larger than the tangential flux, while the difference between both components reduces in the small concave intrusions, where curvature becomes remarkably high.}



\dmmrev{By comparing the TC method with the reference DC solutions, it can be seen the TC model accurately predicts the net contribution of the fluxes in both the normal and tangential directions, even when the tangential fluxes are not negligible.} The differences are mainly confined to two areas: the highly curved concave regions (the far right of the scatter plot), where the TC model tends to be less populated, and the most convex parts (the far left of the scatter plot), where the TC model tends to predict slightly more wrinkled flame fronts.

These regions are expected to have an impact on the behaviour of the flame and the discrepancy between models is mainly associated to a weakening of the flamelet hypothesis \ep{since the tangential fluxes \dmmrev{start influencing} the flame structure}. In these results, the differences are aligned with other tabulated methods that incorporate preferential diffusion~\cite{Bottler2023}. It is worth mentioning that the scattering tends to be more similar between both models as time evolves. 
\dmmrev{As both models may give different responses for the same wavelength (see \ref{subsec:results2}), the response at early time may not be representative of the fully developed flame due to the influence of the original perturbation. The correlation at this early stage actually depends on the initial condition and how the different wavelengths are initially excited.}
\dmmrev{In contrast, once the flame front is fully developed, the influence of the initial conditions dissipates. If a sufficiently large sample of points is taken (i.e., a long enough flame front arc length), it is expected that the points will span all possible states permitted by the model, reflecting the dynamic balance between its stable and unstable responses to perturbations. Therefore,}
it is more reliable and accurate to make comparisons for advanced times.

Overall, the TC model demonstrates good agreement with the reference solution based on detailed chemistry DC. Despite some discrepancies, the tabulated flamelet model effectively captures the flame structure and \ep{flame front distortion,} 
while it offers a significant reduction in computational cost. \ep{In the following, insights into the flame dynamics are given through the analysis of the flame evolution in the linear and non-linear regimes.}


\subsection{Analysis of the linear regime \label{subsec:results2}} \addvspace{10pt}

In this section, the linear \ep{regime is} 
analysed in terms of the dispersion relation, which describes the \ep{initial growth rate of the amplitude $\omega$ triggered by a small perturbation of the flame front \dmmrev{for a given} wavelength ($A(t)=A_0 \, e^{\omega t}$). \dmmrev{The dispersion relation describes}
the flame response for all the wavelengths and gathers the hydrodynamic and thermodiffusive contributions, so it provides very valuable information on the flame behaviour.}


After the initial perturbation, the flame starts propagating, and for the large wavelengths, the perturbation tends to increase and become unstable due to the thermal expansion (Darrieus-Landau instability). Instead, \dmmrev{the flame is 
curved for small wavelengths} modifying the flame speed \ep{due to thermodiffusive effects which may compensate or enhance the hydrodynamic instability (depending on the Lewis number). Finally, \dmmrev{the stabilizing thermodiffusive contribution becomes dominant as the wavelength is reduced, and the perturbation is consequently dissipated leading to 
a stable regime.}} There exists, therefore, a critical wavelength ($\lambda_{crit} \neq 0$) at which the flame is marginally stable \ep{($\omega=0$)}. \ep{Given an equivalence ratio, if the effective Lewis number \emf{$Le_{\mathrm{eff}}$} is smaller than 1, the preferential diffusion contribution tends to destabilize the flame, while if $\emf{Le_{\mathrm{eff}}}<Le_c=1-2/Ze$, with $Ze$ being the Zeldovich number, the net thermodiffusive effects (total contribution of thermal plus species diffusion) are destabilizing. In the case of hydrogen, \dmmrev{destabilizing effects are observed for equivalence ratios smaller than $\phi < 0.75$, showing that the flames considered here ($\phi=0.5$) are subjected to strong thermodiffusive instabilities.}}


For this analysis, the computational setup of Section~\ref{subsubsec: methodology-linear} \ep{together with an initially flame front perturbed according to the}
periodic function $F\left(x\right)=A_0 \mathrm{cos}(k x)$ is used \cite{Berger2022IntrinsicInstabilitiesPremixed_I, Bottler2023}, \ep{where $A_{0}=0.04l_{F}$}. 
\tg{\dmmrev{Isolines of temperature at $T=1000K$ are considered for Cases I and III, while Case II uses $T=1500K$}} to sample the amplitude $A(t)$ 
and calculate the growth rate $\omega$ from eq.~\eqref{eq:growth_rate} until the non-linear regime is achieved.
The values of the wavelengths are selected based on a linear equally spaced discretization of $k l_{F}$ between $0.25$ and $3.5$ with $14$ points. The point at $k=0$ is filled with \dmm{the} value $\omega=0$.

\begin{figure*}
\centering
\includegraphics[width=0.99\textwidth]{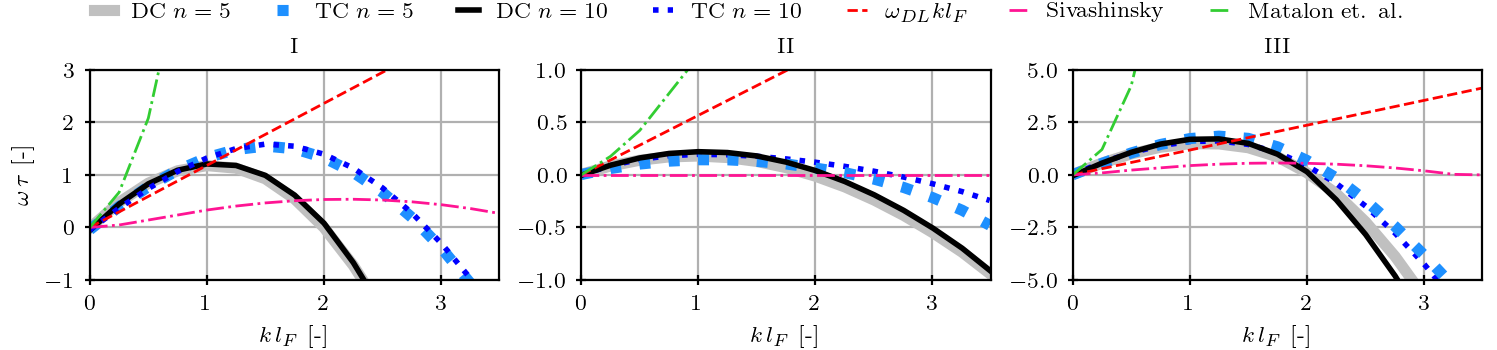}
\caption{\footnotesize Dispersion relation for \ep{Cases I (left), II (center) and III (right)}
for all combinations of model and mesh resolutions. \ep{The relations for the different models are represented with different lines:} 
DC with fine mesh (solid thin black lines), DC with coarse mesh (solid thick silver lines), TC with fine mesh (dotted thin blue lines) and TC with coarse mesh (dotted thick light blue lines). The wave number of the hydrodynamic instability (red dashed line), the model of Matalon et al. from eq.\eqref{eq:matalon} (green dash dotted lines) and the model of Sivashinsky eq.\eqref{eq:sivashinsky} (pink dash dotted lines) are also included for comparison.
}\vspace{-2.5mm}
\label{fig: dispersion_relation}
\end{figure*}

The results of the dispersion relation for the three conditions \ep{and different mesh resolutions} are shown in Fig.~\ref{fig: dispersion_relation}. 
The TC method predicts less stable flame fronts featuring higher growth rates \ep{and larger unstable regions}, though this overprediction \ep{is notably moderated} 
for Cases II and III. Such higher growths contribute to a more wrinkled flame front as observed in Fig.~\ref{fig: contours}, especially for Case I. This observation is consistent with a similar study 
based on the flamelet concept, but using different flow conditions~\cite{Bottler2023}.
Moreover, differences between the TC and DC models tend to be magnified at high wavenumbers. Such divergence is attributed to a weakening in the flamelet hypothesis \tg{of 
thin flames} for highly curved flame fronts (large wavenumbers)  where a complex flame structure deviates from the manifold constructed from \ep{the unstretched} one-dimensional flames \ep{due to the 
\dmmrev{influence of the tangential fluxes in largely curved regions,}
as show in Fig. \ref{fig: diffusive fluxes}}. 
Conversely, for cases at higher temperature (Case II) or pressure (Case III), the proposed model predicts more extensive ranges of wavenumbers, and indeed, only differences are detected in the stable range of wavenumbers ($\omega <0$). The origin for this improvement is attributed to a) an attenuation of the thermodiffusive instability together with more moderate hydrodynamic instabilities \dmmrev{for the high temperature \ep{(Case II)},} which entail a less corrugated and more flattened flame front, and b) a reinforcement of the flamelet hypothesis \ep{(higher Zeldovich number)} \dmmrev{in the case of higher pressure \ep{(Case III)},}
which compensates the increased \ep{flame} wrinkling.

From a practical point of view, the \ep{satisfactory} performance of the model 
for high temperature and pressure offers high potential of this method to be applied at relevant engine conditions \dmmrev{with} reduced computational cost.
In addition, it is essential to emphasise the capability of the proposed model to recover more extensive ranges of wavenumbers under such conditions, as observed for Cases II and III, where only differences are detected in the stable range of wavenumbers. Finally, a weak influence of the mesh resolution is observed on the dispersion relation for the considered cell sizes.

To provide a more comprehensive analysis of the \ep{different instabilities,} 
Figure~\ref{fig: dispersion_relation} incorporates the line for the Darrieus-Landau instability 
alongside theoretical models from Matalon et al. \cite{matalon2007intrinsic, Matalon2009, Matalon2018} and Sivashinsky \cite{Sivashinsky1977DiffusionalThermalTheoryOfCellularFlames, Sivashinsky1977NonlinearAnalysisOfHydrodynamicInstability}. The instability arises due to density differences between the unburnt and burnt mixture, characterized by the density ratio $\sigma := \rho_u / \rho_b$ (Darrieus and Landau~\cite{Darrieus1945, LANDAU1988403}). 
This instability is characterized by a \ep{normalised} growth rate, $\omega_{DL}$ \ep{($\omega \tau = \omega_{DL} s_L k $)}, defined by the following equation:
\begin{equation}\label{eq:darrius-landau}
    \omega_{DL} = \frac{\sqrt{ \sigma^{3} + \sigma^{2} - \sigma} - \sigma}{\sigma + 1}.
\end{equation}
Notably, the dispersion relation line \ep{due to the hydrodynamic instability} 
$\omega \tau = \omega_{DL} k l_{F}$ is always positive for any wavenumber ($\sigma>1$),\ep{ that is,} 
the hydrodynamic instability 
destabilizes the flame \ep{for every wave number.}

Matalon et al.~\cite{Matalon2009, Matalon2018} proposed a wavelength model for premixed flames based on the hydrodynamic theory using a multi-scale approach and considering the influence of the flame speed with the flame stretch.
This formulation incorporates preferential diffusion effects and is expressed up to a second-order approximation by the following equation:

\begin{equation}\label{eq:matalon}
    \omega_{M} \tau = \omega_{DL} k l_{F} - \frac{l_{D}}{l_{F}} \left[B_{1} + Ze \left(Le_{\mathrm{eff}} - 1\right)B_2 + Pr \, B_3\right] (k l_F)^2.
\end{equation}


\noindent In this equation, $l_{D}$ represents the diffusive flame thickness, defined as the ratio of thermal diffusivity $D_{th} = \lambda / \rho c_{p}$ to the laminar flame speed, $l_{D} = D_{th}/s_L$. The Zeldovich number is denoted by $Ze$, the Prandtl number by $Pr$, and $Le_{\mathrm{eff}}$ is the effective Lewis number used in the model. Coefficients $B_1$, $B_2$, and $B_3$ are positive constants that depend on the ratio of thermal conductivity to temperature, scaled with respect to the unburnt mixture values. Details on the calculation of $B_1$, $B_2$, and $B_3$ are provided in 
~\ref{appendix:disp_rel-matalon}, while the calculations for $Le_{\mathrm{eff}}$ and $Ze$ are detailed in 
\ref{appendix:effective-Lewis} and~\ref{appendix:Zeldovich-number}, respectively.


The Sivashinsky dispersion relation $\omega_{S}(k)$ is obtained by solving the following equation:

\begin{equation}\label{eq:sivashinsky}
    \frac{\left[Le_{\mathrm{eff}}-q(\omega_{S}, k)\right]\left[p(\omega_{S},k)-r(\omega_{S},k)\right]}{Le_{\mathrm{eff}}-q(\omega_{S},k)+p(\omega_{S},k)-1}-\frac{Ze}{2}=0,
\end{equation}
where the terms $p(\omega_S, k)$, $q(\omega_S, k)$, and $r(\omega_S, k)$ are defined in ~\ref{appendix:disp_rel-sivashinsky}. Their explicit dependence on the expansion rate and wave number, $\omega_S$ and $k$, is shown for clarity.

\ep{On the one hand, it is observed that both DC and TC models give a \dmmrev{response above the predicted} Darrieus-Landau model (hydrodynamic instability) for low wave numbers ($k\,l_F$ in the range 0 to 1-1.5) \dmmrev{for Cases I and III.}
\tg{This is due to the significant destabilizing effect \dmmrev{caused by the low Lewis number of hydrogen,} which influences even the lower wave numbers where hydrodynamic instability prevails.} 
Such behaviour is also predicted by Matalon et al.~\cite{Matalon2009, Matalon2018} model, despite it largely overestimates the growth rates. In a similar manner, Sivashinsky model gives positive growth rates for a large range of wave numbers caused by the thermodiffusive effects, showing that the non-unity Lewis number effects have a strong destabilizing component. On the other hand, the thermodiffusive destabilizing effects are remarkably attenuated in Case II as expected from the comparison of the effective Lewis number \emf{$Le_{\mathrm{eff}}$} 0.34 (see \ref{appendix:effective-Lewis}) and the critical Lewis number $Le_c=1-2/Ze=0.41$. \dmmrev{In these conditions,} the stabilizing thermodiffusive effects caused by the flame front curvature and the non-unity Lewis effects are cancelled \cite{Sivashinsky1977DiffusionalThermalTheoryOfCellularFlames}. In this case, both DC and TC models show a dispersion relation below the Darrieus-Landau line for all the wave numbers with a slope at $\omega=0$, which approximates well to $\omega_{DL}$. In line with this behaviour, Sivashinsky model predicts a marginal stable response of the flame, while Matalon et al. predicts a more moderate behaviour than Cases I and III \dmmrev{but with a non-negligible} overestimation of the growth rates.}


\ep{In summary,} the analysis of the linear regime shows good correlation for TC with DC, though some disagreements \ep{tend to be observed at high wave numbers.}
The TC model can quantitatively capture, \ep{however,} the flame response specially for the Cases II and III, \ep{that is, when increasing either the temperature or the pressure. This fact has 
\dmmrev{some practical implications
since such conditions are 
more similar to engine operation.}}
\ep{To close the analysis, the non-linear regime is analysed in the following subsection through the evaluation of the} 
macroscopic flame quantities.

\subsection{Analysis of the non-linear regime - Macroscopic flame parameters}\label{subsec:results3}

Previous subsections have shown an important effect
of the thermodiffusive fluxes on the flame \ep{structure and its} dynamics in the linear \ep{regime.} 
In this subsection, an evaluation of \ep{relevant} global flame parameters for the three operating conditions is conducted \dmmrev{in the non-linear regime, that is, once the flame has fully developed.}
\ep{Such parameters comprise the consumption speed, the flame surface area and \dmmrev{the reactivity ratio as described by Berger et al.}~\cite{Berger2022IntrinsicInstabilitiesPremixed_II}.}

A relevant parameter that influences the design of practical systems is the consumption speed $s_{c}$, which measures the fuel consumption rate and plays a fundamental role in the burning velocity of the flame front. It is defined by integrating the fuel reaction rate over the domain \dmm{and can be obtained by}:

\begin{equation}\label{eq:consumption_speed} s_{c} = - \frac{1}{\rho_{u} Y_{{H_2},u} L_{x}} \int \dot{\omega}_{H_{2}} \,\, dxdy,\end{equation}

\noindent where $\rho_u$ and $Y_{{H_2},u}$ represent the density and hydrogen mass fraction on the unburnt side, respectively.

\begin{figure*}
\centering
\vspace{-0.4 in}
\includegraphics[width=0.99\linewidth]{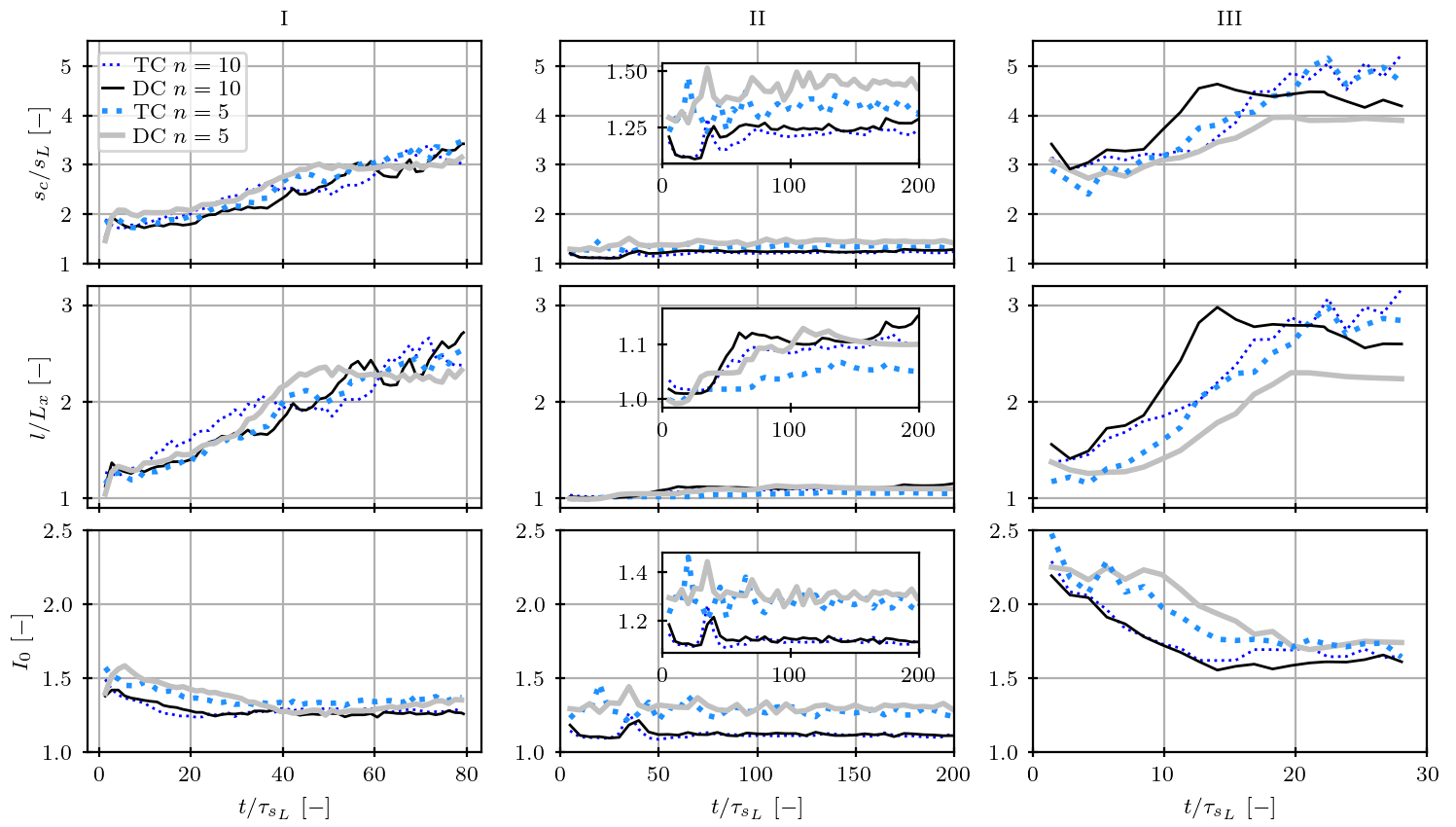}
\vspace{10 pt}
\caption{\footnotesize
Macroscopic quantities for the simulations of Table.~\ref{tab:main_cases} for model DC with $n=10$ (black thin solid lines), DC with $n=5$ (silver coarse solid lines), TC with $n=10$ (blue thin dotted lines) and TC with $n=5$ (light blue coarse dotted lines). Results of the consumption speed ratio $s_{c}/s_{L}$ (first row), the flame surface ratio $l/L_{x}$ (second row) and the reactivity ratio $I_{0}$ (third row) for \ep{Case I (left column), II (central column) and III (right column).}
}\vspace{-2.5mm}
\label{fig: speed}
\end{figure*}

Another parameter of interest is the flame surface area \dmm{$l$} defined as:
\begin{equation}\label{eq:surface_area}
    l=\int |\vec{\nabla} c| dxdy.
\end{equation}
The flame surface \ep{area} measures the area of the flame front and is linked to the consumption speed via the displacement speed $s_{d}$, whose relation is given by:
\begin{equation}
    s_{c} = \frac{l}{\rho_{u} Y_{H_2, u} L_{x}} \langle \rho s_{d}\rangle,
\end{equation}
and $\langle \rho s_{d} \rangle =\frac{\int_{\Omega}  \rho s_{d}|\nabla C| d V}{\int_{\Omega}|\nabla C| d V}$.
Therefore, \dmm{a complete characterisation of these macroscopic quantities can be made by using $l$ and $s_{c}$. Another quantity of interest is the displacement speed $s_{d}$, which was studied by Chu et al.~\cite{Chu2023EffectsDifferentialDiffusion}.
By comparing these quantities with those of 
the unstretched flames $s_{c}/s_{L}$, the reactivity ratio $I_{0}$ can be obtained as}:
\begin{equation}\label{eq:reactivity} I_{0} = \frac{s_{c}}{s_{L}} \frac{L_x}{l}, \end{equation}

\noindent where the ratio $l/L_{x}$ is referred to as the wrinkling factor and measures the area of the flame front with respect to the reference domain length. The reactivity ratio quantifies the deviation from the unstretched one-dimensional flame of the feeding mixture.
A representation of these three quantities for the different operating conditions and mesh resolutions is shown in Fig.~\ref{fig: speed}.
Two regimes can be identified to describe the dynamics of the flame. 
In the first part of the evolution, both the flame \ep{surface area} 
and the consumption speed increase rapidly (mainly for Cases I and III) due to the wrinkling of the flame front after the linear regime. This part is followed by a steady regime with oscillations around constant values for $l/L_{x}$ and $s_c$. The reactivity ratio $I_{0}$ slightly reduces in the first phase, followed by a steady-state plateau that characterises the second part.
In contrast to Cases I and III, 
Case II does not exhibit such behaviours, since the instabilities are weak at high temperature and play a minor role on the global burning rates.
Indeed, the consumption speed in this case reaches \ep{the} steady state faster and features lower enhancements of flame speed than \ep{for} the rest of \ep{the} cases.
The reactivity ratio is low \ep{compared to the other cases, therefore, showing \dmmrev{a flame with} a more similar behaviour to the unstretched one-dimensional flame as a consequence of the less wrinkled flame front (lower curvatures).}

The results show \ep{that} the TC model reproduces well the mean values and fluctuations obtained by DC independently of the conditions. Moreover, the stabilisation time for the reactivity ratio is also predicted accurately. The mesh resolution leads to a slight over-prediction of the steady state for the reactivity ratio, but overall, the agreement is quite remarkable. 

The consumption speed and reactivity factor increase when the instabilities are promoted (increasing pressure, decreasing temperature), see Fig.~\ref{fig: speed}. The reactivity ratio $I_0$ is proportional to the product of the consumption speed and the \ep{flame surface area,} 
as described by Eq.~\eqref{eq:reactivity}. This dependence arises from the separation between the thermochemical states of the actual flame and the unstretched one-dimensional flame, in agreement with the results shown in Fig.~\ref{fig: histograms} and those from  literature~\cite{Berger2022IntrinsicInstabilitiesPremixed_II}. Similarly, the flame \ep{front arc } length increases, even in the case of increasing pressures, though a highly insensitive dependence was reported in~\cite{Berger2022IntrinsicInstabilitiesPremixed_II}. It is essential to highlight that Soret effect was included in that work, but not in the present model. Nevertheless, these results and major conclusions are consistent and aligned with previous observations~\cite{Berger2022IntrinsicInstabilitiesPremixed_II,LAPENNA2024113126}.

\begin{figure*}
\centering
\vspace{-0.4 in}
\includegraphics[width=0.99\linewidth]{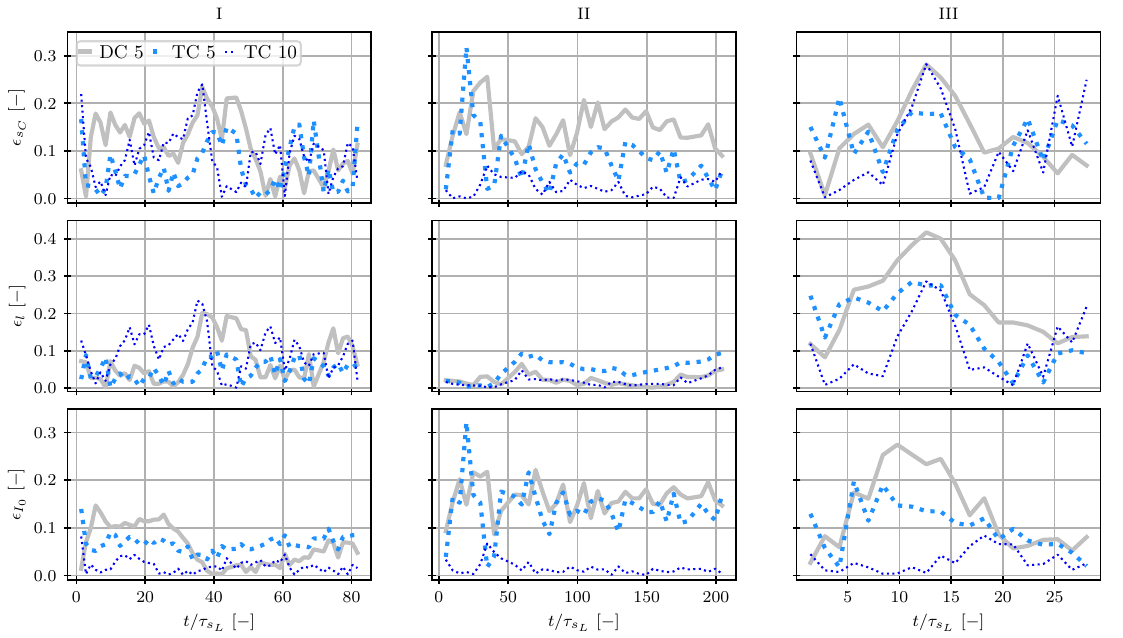}
\vspace{10 pt}
\caption{\footnotesize
Relative error of macroscopic quantities, $Q$, with respect to the simulations for model DC with $n=10$ ($\epsilon = |Q^{DC,10} - Q^{model, n} / Q^{DC,10}|$). DC with $n=5$ (silver coarse solid lines), TC with $n=5$ (light blue coarse dotted lines) and TC with $n=10$ (blue thin dotted lines). Results of the consumption speed ratio $s_{C}/s_{L}$ ($\epsilon_{s_{C}}$, first row), the flame surface ratio $l/L_{x}$ ($\epsilon_{l}$, second row) and the reactivity ratio $I_{0}$ ($\epsilon_{I_{0}}$, third row) for each Case I, II and III in each column 1, 2 and 3, respectively.}
\label{fig: flame speed error}
\end{figure*}

Finally, the \ep{temporal} 
evolution of the error \ep{related to the prediction of these global parameters is} 
shown in Fig.~\ref{fig: flame speed error}.
The reference solution for each condition 
is obtained using the detailed chemistry solver DC with  $n=10$ points per $l_{F}$.
Thus, the relative error for the other simulations, labelled as (model, $n$), is $\epsilon = |\frac{s^{DC,10}_{C} - s^{model, n}_{C}}{s^{DC,10}_{C}}|$.

The results presented in Fig.~\ref{fig: flame speed error} confirm the predictive capabilities of the tabulated approach to recover the consumption speed and the rest of quantities. The oscillating errors are kept below 10\% except for some periods that could reach peaks up to 20\%. Nevertheless, these errors are rapidly reduced and remain below 10\% afterwards. This is more pronounced for the case with finer mesh, where the TC method excels at capturing the ratio $I_{0}$, where the error remains essentially constant at 5\%. This means that while the consumption speed and area errors may increase compared to the DC solution, the $I_0$ value does not deviate from the correct solution, resulting in consistent and reliable results. \ep{Moreover, the reduction of the hydrodynamic and thermodiffusive instabilities for the high temperature Case II is also translated into a decrease of the errors for TC with $n=10$.}

In light of the results, it can be stated that the proposed TC model can accurately capture the flame macroscopic characteristics with small deviations with respect to the DC model in agreement with other tabulated approaches \cite{Bottler2023}. 
Finally, it can be concluded that the proposed flamelet tabulated method is sensitive to the mesh resolution, though its influence on the results is limited.

Considering the prediction of the dispersion relations in the linear regime (see Fig.~\ref{fig: dispersion_relation}), differences in the global flame descriptors are attributed to the effect of the small wavelengths. 
These wavenumbers are shown to play a fundamental role, 
especially for the highly corrugated flame fronts (Cases I and III). 

\ep{It is worth mentioning that while} the dispersion relation \ep{predicts the flame response for each wave number after a short period of time, it} partially overlooks the \ep{real response of the flame submitted to strong preferential diffusion since the} variability of the mixture fraction field downstream of the flame front due to the wrinkling \ep{was not included when obtaining the dispersion relations (mixture leaning at concave regions, enrichment at convex ones). This is a source of discrepancies whose effect was not appreciated in the linear regime analysis and can introduce some deviations between models in the non-linear regime. In addition,} 
the mesh resolution may also affect such \ep{mixture fraction} variability by introducing numerical errors that are considered a second order source of discrepancies. 
Nevertheless, considering that these configurations are challenging due to their rich physical content, it can be concluded that the proposed flamelet tabulation model features powerful predictive capabilities and is suitable for capturing the phenomena involved.

\section{Conclusions}\label{sec: conclusions}

This study presents a numerical investigation of the
capabilities of a tabulated flamelet model with mixture-averaged diffusion \ep{transport} to predict thermodiffusive instabilities in lean hydrogen premixed flames. The model incorporates the effects of preferential diffusion and velocity corrections, resulting in a complete formulation that accounts for the cross-interactions between the controlling variables. This approach is compared with reference solutions based on finite rate chemistry with direct integration using the same transport model. The test cases include three different operating conditions that account for variations in temperature and pressure to replicate more realistic engine conditions. Results with two different mesh spacing are compared to evaluate the influence of mesh resolution on the predictive capabilities of the model.

The results show that the model can capture the wrinkling of the flame front for different conditions and the formation of cusps and cellular structures promoted by the thermodiffusive effects, which agree well with theoretical results and \dmm{previous} numerical studies. A more quantitative comparison through the analysis of the dispersion relation \dmm{shows a tendency of the model to 
overpredict the growth \dmmrev{of} the perturbation 
and, therefore, an overprediction of the} flame front wrinkling. The differences tend to be reduced when the temperature or pressure increases, \ep{that is, when either the flame instabilities are weakened }\tg{ or 
when the flamelet hypothesis is recovered}. Moreover, a limited influence of the mesh \dmm{resolution} is observed in the linear regime. 
The results also keep the regions of hydrodynamic and thermodiffusive instability consistent, \dmm{revealing a strong potential 
of this approach to solve complex physical problems at engine-relevant conditions due to its computational efficiency and reduced computational cost.}

When extending the analysis to the non-linear regime, it has been shown that the model can capture most of the flame global parameters with errors bounded to \dmm{10\% for highly corrugated flame fronts, with some local errors reaching up to 20\% in 
\ep{some of the } conditions that are, \ep{however,} reduced after the initialization}. In contrast, this error is reduced for weakly wrinkled flames for the tested conditions. In consequence, it is concluded that the small wavelengths have a non-negligible but limited influence on the characterization of the flame, and the current formulation shows excellent potential to describe the flame evolution in thermodiffusively unstable conditions, \dmm{making this model suitable for its application to practical configurations, including its extension to large-eddy simulations.}

\vspace{2mm}

\textbf{Declaration of competing interest}: The authors declare that they have no known competing financial interests or personal relationships that could have appeared to influence the work reported in this paper.

\section{Acknowledgements}

The research leading to these results has received funding from the European Union's Horizon 2020 Programme under the CoEC project, grant agreement No. 952181, the HyInHeat project grant agreement No. 101091456 and the H2AERO CPP2022-
009921 project from the Ministerio de Ciencia e Innovación.
EMF acknowledges the grant Ajuts Joan Oró (AGAUR) per a la contractació de personal investigador predoctoral en formació (FI 2023) cofinanced by the EU, Generalitat de Catalunya: Departament de Recerca i Universitats and Agencia de Gestión de Ayudas Universitarias y de Investigación. DM acknowledges the grant Ramón y Cajal RYC2021-034654-I.

\clearpage

\appendix

\section{Scatter plots for radical species}\label{appendix:radicals}

\begin{figure}
    \centering
    \includegraphics[width=1.0\textwidth]{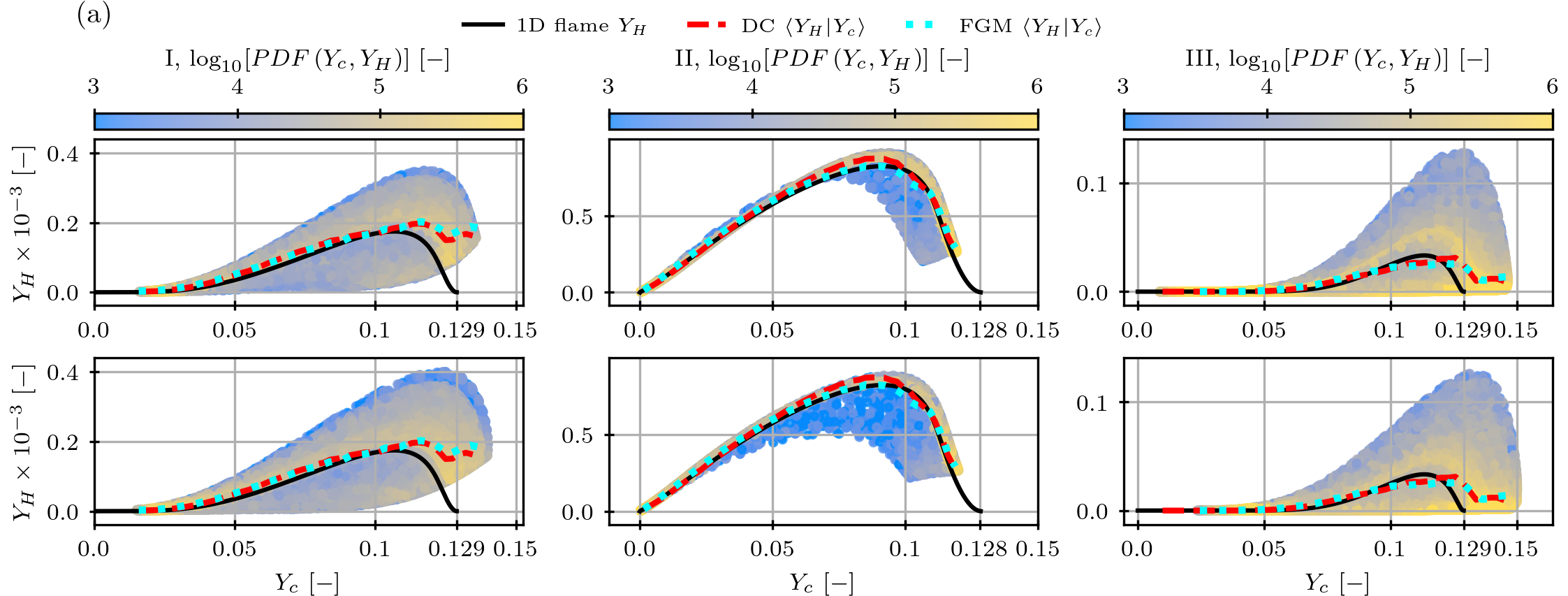}
    \includegraphics[width=1.0\textwidth]{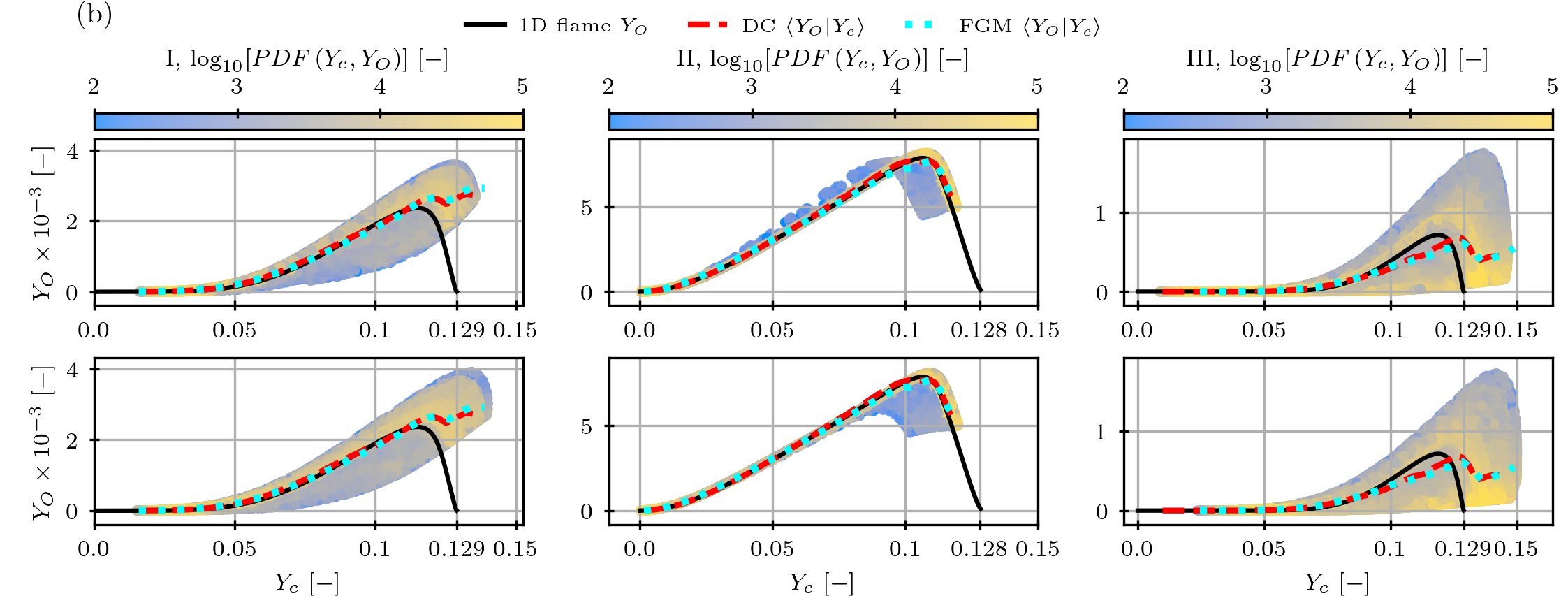}
    \includegraphics[width=1.0\textwidth]{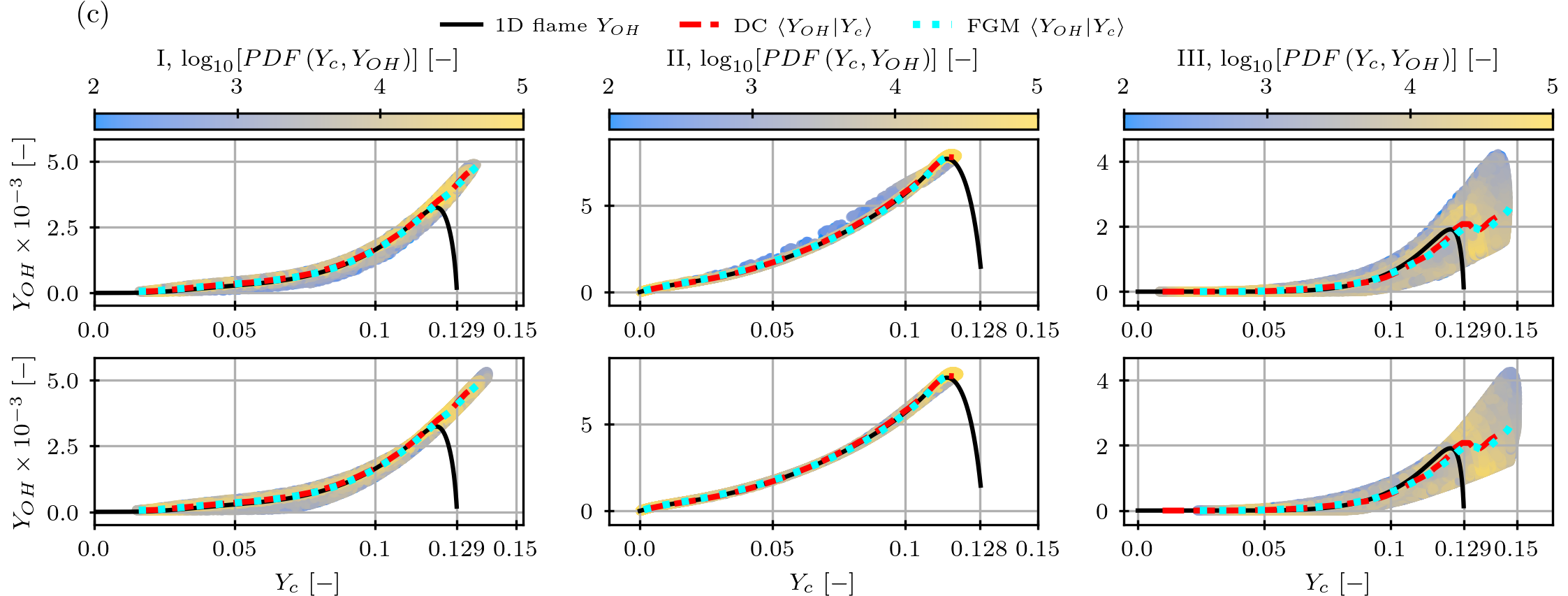}
    \caption{Each panel presents scatter plots of the mass fraction of a radical species versus $Y_c$, coloured by the $\mathrm{log}_{10}$ values of their joint PDFs (jPDFs), for flames I (left column), II (central column), and III (right colum) using the models DC (top) and TC (bottom) with a mesh resolution of $n=10$ at the time instants shown in Fig.~\ref{fig: contours}. The plots also include the values for the one-dimensional unstretched flame at the inlet mixture fraction (black solid lines) and the conditional averages for each model (red solid lines for DC and cyan dotted lines for TC) with $n=10$, for reference. \textbf{(a)}: $Y_{\mathrm{H}}$; \textbf{(b)}: $Y_{\mathrm{O}}$; \textbf{(c)}: $Y_{\mathrm{OH}}$.}
    \label{fig:radicals}
\end{figure}

\emf{This appendix presents the results of the joint PDF distributions of the reactive progress variable $Y_{c}$ and the mass fractions of the three radicals: $Y_{\mathrm{H}}$, $Y_{\mathrm{O}}$, and $Y_{\mathrm{OH}}$ as shown in the different panels of Fig.~\ref{fig:radicals}.}




\section{Theoretical dispersion relations quantities}\label{appendix:disp_rel}

\subsection{Matalon dispersion relationship coefficients}\label{appendix:disp_rel-matalon}

The coefficients $B_1$, $B_2$ and $B_3$ for the dispersion relationship from Matalon et al. of eq.~\eqref{eq:matalon} are defined as:

\begin{equation}\label{eq:B1}
    B_1 = \frac{\sigma / 2}{\sigma+(\sigma+1) \omega_{\mathrm{DL}}}\left\{\frac{\sigma\left(2 \omega_{\mathrm{DL}}+\sigma+1\right)}{\sigma-1} \int_1^\sigma \frac{\tilde{\lambda}(x)}{x} \mathrm{~d} x+\int_1^\sigma \tilde{\lambda}(x) \mathrm{d} x\right\},
\end{equation}

\begin{equation}\label{eq:B2}
    B_2 = \frac{\sigma / 2}{\sigma+(\sigma+1) \omega_{\mathrm{DL}}}\left\{\frac{\left(1+\omega_{\mathrm{DL}}\right)\left(\sigma+\omega_{\mathrm{DL}}\right)}{\sigma-1} \int_1^\sigma \ln \left(\frac{\sigma-1}{x-1}\right) \frac{\tilde{\lambda}(x)}{x} \mathrm{~d} x\right\},
\end{equation}

\begin{equation}\label{eq:B3}
    B_3 = \frac{\sigma}{\sigma+(\sigma+1) \omega_{\mathrm{DL}}}\left\{(\sigma-1) \tilde{\lambda}(\sigma)-\int_1^\sigma \tilde{\lambda}(x) \mathrm{d} x\right\},
\end{equation}

\noindent where $x$ is the scaled temperature $x:=T/T_u$ and $\tilde{\lambda}$ is the scaled thermal conductivity $\tilde{\lambda}(x):=\lambda / \lambda_{u}$, \ep{both normalized with the values at the unburnt gases.} 
The values in the integral are calculated by numerical integration of the one dimensional unstretched flames. The values for $\sigma$ and the $B$-coefficients are shown at columns 2 to 5 of Table~\ref{tab:appendix}. 

\subsection{Sivashinsky dispersion relationship equations}\label{appendix:disp_rel-sivashinsky}

The implicit dispersion relation from Sivashisnky's eq.~\eqref{eq:sivashinsky} quantities are defined as

\begin{equation}\label{eq:p}
p=\frac{1}{2}\left[1+\sqrt{1+4\left(l_D \omega / s_L + l_D^2 k^2\right)}\right],
\end{equation}

\begin{equation}\label{eq:q}
q=\frac{Le_{\mathrm{eff}}}{2}\left[1+\sqrt{1+\frac{4\left(l_D \omega / s_L Le_{\mathrm{eff}}+l_D^2 k^2\right)}{Le_{\mathrm{eff}}^2}}\right],
\end{equation}

\begin{equation}\label{eq:r}
r=\frac{1}{2}\left[1-\sqrt{1+4\left(l_D \omega / s_L+l_D^2 k^2\right)}\right] .
\end{equation}

\subsection{Calculation of effective Lewis number}\label{appendix:effective-Lewis}

To calculate the effective Lewis number required for the Matalon et al.  \cite{Matalon2018} and \ep{Sivashisnky dispersion relation}, we use the formula originally defined by \cite{Joulin1981LinearStability} and later simplified by \cite{Berger2022IntrinsicInstabilitiesPremixed_I}, as follows:

\begin{equation}\label{eq:effective-Lewis}
    L e_{\mathrm{eff}} = 1 + \frac{\left(L e_{\mathrm{E}} - 1\right) + \left(L e_{\mathrm{D}} - 1\right) \mathcal{A}}{1 + \mathcal{A}},
\end{equation}

\noindent where $\mathcal{A} = 1 + Ze \left(\phi^{-1} - 1\right)$, $L e_{\mathrm{E}}$ refers to the Lewis number of the excess reactant, and $L e_{\mathrm{D}}$ refers to the Lewis number of the deficient reactant. The Lewis numbers for these reactants are obtained from the burnt value of the one-dimensional flame. The resulting effective Lewis numbers are presented in Appendix Table~\ref{tab:appendix}, which align with the results of \cite{Berger2022IntrinsicInstabilitiesPremixed_I}.

\subsection{Calculation of Zeldovich number}\label{appendix:Zeldovich-number}

The Zeldovich number is a dimensionless defined by the equation: 
\begin{equation}\label{eq:Zeldovich}
    Ze = \frac{E_a}{R} \frac{T_b - T_u}{T_b^2} = 2 (T_b - T_u)  \frac{d\,(\mathrm{ln}(\rho_u s_L))} {d\,T_b},
\end{equation}

\noindent where $E_a$ is the activation energy \cite{Peters1987TheAsymptoticStructure} and $R$ is the universal gas constant. As seen from eq.~\eqref{eq:Zeldovich}, the activation energy is related to the calculation of a derivative. This derivative is taken using the methodology from \cite{Sun1999DynamicsOfWeaklyStretchedFlames}. The method involves generating two additional one-dimensional unstretched flame solutions. These solutions use a diluted mixture with a very small amount ($\pm$0.3\%) of extra nitrogen mass fraction, while keeping the equivalence ratio, pressure, and unburnt temperature fixed. The Zeldovich numbers for different cases are presented in the last column of Table~\ref{tab:appendix}. 

\begin{table}
\centering
\begin{center}
\begin{tabular}{|c|c|c|c|c|c|c|c|c|}
\hline
Case  & $\sigma$ $[-]$ & $B1$ & $B2$ & $B3$ & $Le_{\mathrm{eff}}$ $[-]$& $Pr$ $[-]$ & $l_{D}$ $[\mu m]$ & $Ze$ $[-]$\\ \hline
I & 5.03 & 8.04 & 2.04 & 1.19 & 0.36 & 0.560 & 67.4 & 9.2\\ 
II & 2.58 & 3.79 & 1.20 & 0.11 & 0.34 & 0.572 & 26.3 & 3.4\\ 
III & 5.03 & 7.87 & 2.01 & 1.26 & 0.44 & 0.560 & 29.8 & 12.9\\ 
\hline
\end{tabular}
\end{center}
\caption{Flame properties of the corresponding unstretched one-dimensional flames required for the calculation of theoretical dispersion relationship from Matalon et al. of eq.~\eqref{eq:matalon} and Sivashinsky eq.~\eqref{eq:sivashinsky}.}
\vspace{-2mm}
\label{tab:appendix}
\end{table}

\bibliographystyle{elsarticle-num}
\bibliography{biblio}

\end{document}